%% file: ms.tex
\documentclass[fleqn,10pt]{wlscirep}
\usepackage[utf8]{inputenc}
\usepackage[T1]{fontenc}
\usepackage{graphicx}
\usepackage{listings}
\usepackage{ulem}
\usepackage{mathtools}
\usepackage{xcolor}
\usepackage{verbatim}
\DeclareSymbolFont{boldoperators}{OT1}{cmr}{bx}{n}
\SetSymbolFont{boldoperators}{bold}{OT1}{cmr}{bx}{n}
\edef\bar{\unexpanded{\protect\mathaccentV{bar}}\number\symboldoperators16}

\usepackage[font=small,labelfont=bf, figurename=Fig.]{caption}

\usepackage{makecell}
\definecolor{ngreen}{rgb}{0,0.6,0}
\newcommand{\grn}{\color{ngreen}}
\definecolor{nred}{rgb}{0.8,0,0}
\newcommand{\redn}{\color{nred}}

\title{A convolutional neural network for defect classification in Bragg coherent X-ray diffraction}

\author[1,2,+]{Bruce Lim}
\author[2,+]{Ewen Bellec}
\author[2,3,*,+]{Maxime Dupraz}
\author[2]{Steven Leake}
\author[4]{Andrea Resta}
\author[4]{Alessandro Coati}
\author[5]{Michael Sprung}
\author[6]{Ehud Almog}
\author[6]{Eugen Rabkin}
\author[2]{Tobias Schulli}
\author[2,3,*]{Marie-Ingrid Richard}
\affil[1]{Univ. Grenoble Alpes, Grenoble INP-Phelma, 3 Parvis Louis Néel, BP 257, 38016 Grenoble, France}
\affil[2]{Univ. Grenoble Alpes, CEA Grenoble, IRIG, MEM, NRS, 17 rue des Martyrs 38000 Grenoble, France}
\affil[3]{ESRF - The European Synchrotron, 71 Avenue des Martyrs, Grenoble 38000, France} 
\affil[4]{Synchrotron SOLEIL, L’Orme des Merisiers, Saint-Aubin, BP48, 91192 Gif-sur-Yvette, France}
\affil[5]{Deutsches Elektronen-Synchrotron (DESY), D-22607 Hamburg, Germany}
\affil[6]{Department of Materials Science and Engineering, Technion-Israel Institute of Technology, 3200003, Haifa, Israel}

\affil[*]{maxime.dupraz@esrf.fr;mrichard@esrf.fr}

\affil[+]{these authors contributed equally to this work}

\begin{abstract}
{
Coherent diffraction imaging enables the imaging of individual defects, such as dislocations or stacking faults, in materials. These defects and their surrounding elastic strain fields have a critical influence on the macroscopic properties and functionality of materials. However, their identification in Bragg coherent diffraction imaging remains a challenge and requires significant data mining. The ability to identify defects from the diffraction pattern alone would be a significant advantage when targeting specific defect types and accelerates experiment design and execution.
Here, we exploit a computational tool based on a three-dimensional (3D) parametric atomistic model and a convolutional neural network to predict dislocations in a crystal from its 3D coherent diffraction pattern. Simulated diffraction patterns from several thousands of relaxed atomistic configurations of nanocrystals are used to train the neural network and to predict the presence or absence of dislocations as well as their type (screw or edge). Our study paves the way for defect recognition in 3D coherent diffraction patterns for material science.

}

\end{abstract}
\begin{document}

\flushbottom
\maketitle
%
%
\thispagestyle{empty}

\section*{Introduction}
Defect detection and classification are important issues in material science, as defects strongly influence the properties of materials \cite{Wang1999,Bei2008,Ohno2008,Bittner2017}. Although metallurgy has long recognized the importance of defects for the macroscopic mechanical properties (\textit{e. g.} such as enhanced yield strength of steel), their more widespread influence in other fields of material science is still lacking detailed understanding. Nevertheless, the concept of strain engineering in a vast variety of functional materials is attracting a lot of attention, opening great opportunities for the design and optimisation of the mechanical, optical, electrical or catalytic properties of materials via deliberate defect manipulation\cite{shin_rational_2013,ulvestad_topological_2015,attariani_defect_2017}. Crystal defects of various nature and length scales are not always adverse but can instead activate specific functionalities, such as improving adsorption affinity or catalytic activity. For instance, twins and stacking faults can improve catalytic efficiency of nanoparticles\cite{Behrens2012} and more generally the strain generated by defects can affect the catalytic activity\cite{vendelbo_visualization_2014}. Similarly, the role of dislocations in battery performance has drawn the attention of scientists and could be a key point for further optimisation\cite{ulvestad_topological_2015}. This defect sensitivity might open new avenues to engineering the properties of nanostructures by introducing specific defects. In order to achieve this goal, it is important to detect and classify defects in nanomaterials to better understand their behaviors (nucleation, propagation, annihilation, defect-defect interaction). 

Unlike perfect crystals that can be described as equilibrium structures, the physics and thermodynamics of defects is much harder to describe with the available theoretical tools. It is thus of greatest relevance to supply imaging techniques capable delivering tomographic reconstructions of the crystal structure in the close environment of defects.
Few experimental techniques can achieve this goal. Among them, transmission electron microscopy (TEM) is routinely used to image dislocations in real space by selecting relevant diffraction vectors, according to established invisibility criteria\cite{carter_transmission_1996}. It has atomic resolution and can directly image individual crystal defects. However, the technique is hindered by several constraints related to sample preparation. These constraints are relaxed for X-rays, which have a great potential to study defects in crystals. With the advent of new generation synchrotron sources with higher coherent flux, a very attractive technique to probe the microstructure of defects has emerged: coherent X-ray diffraction (CXD)\cite{livet_diffraction_2007,sutton_review_2008}. In Bragg geometry, it probes the local deviation from the perfect crystal lattice and is therefore highly sensitive to elastic strain \cite{Beutier2013} and crystal defects such as stacking faults \cite{Favre_Nicolin_2010} or dislocation loops \cite{Jacques2011}. In the past two decades, the technique has been turned into an imaging technique (Bragg Coherent Diffraction Imaging, BCDI), combining measurements of three-dimensional (3D) Coherent X-ray diffraction patterns (CXDPs) with phase retrieval algorithms \cite{Fienup:82,Marchesini2003}, to obtain a spatial reconstruction of isolated nanoscale objects \cite{Robinson2009}. The technique has been used successfully to image the strain field in defective nanocrystals \cite{ulvestad_topological_2015,Dupraz2017} including for relatively complex defect configurations \cite{Clark2015,Hofmann2020}, but tends to fail for highly strained systems. In addition, phase retrieval algorithms are relatively slow, while a live evaluation of the data is often required during \textit{in situ} and \textit{operando} experiments. This is particularly true in the case of Bragg ptychography, which requires a considerable amount of data. There is therefore an interest in understanding CXDPs qualitatively and interpreting them directly in reciprocal space. Depending on their type and on the measured Bragg reflection, single crystal defects have indeed a unique signature on CXDPs which enables their identification directly from the reciprocal space data \cite{dupraz2015signature}. For instance, a screw dislocation will lead to a ring-shaped Bragg diffraction signal, if the Burgers vector \textbf{b} of the dislocation is parallel to the scattering vector at the measured Bragg position, \textbf{g}. 

For identifying defects, pattern classification\cite{krueger_fault_2001,caulier_specific_2008} and neural networks (NN) for fault detection\cite{jueptner_application_1994} have been previously used, for example in diffraction phase microscopy.
Deep learning has also been used successfully for optical surface-defect detection \cite{Nash2018,Tabernik2020,doi:10.1177/1687814018766682} and for defect segmentation in scanning transmission electron microscopy \cite{Roberts2019}.
These methods are therefore relevant to detect and classify defects in CXDPs, which are very sensitive to the defect type. The need of extensive training sets and prior data with different type of defects is one of the main difficulties to overcome with these computational methods. These requirements could potentially limit their performance and practical feasibility. However, with the exponential advancements in computational resources\cite{denning_exponential_2016} and the possibility of ultra-fast atomistic relaxation and computation of diffraction patterns with massive parallelism or graphical processing units (GPUs), it is now straightforward to calculate the 3D CXDPs of single nanocrystals from their atomistic configurations. These configurations can be generated by varying the type and location of the crystal defects and then relaxed by energy minimization. The relaxation of the faulted crystal structure allows to model accurately the crystal defect and has been shown to have a large impact on CXDPs\cite{dupraz2015signature}, leading to a better agreement between the simulated 3D CXDPs and experimental measurements. 

While models have been widely applied to generate 2D images, generation of 3D structures is a nascent field. For example, a deep learning NN model has been recently successfully developed for classification of crystal structures from 2D diffraction maps of more than 100,000 simulated crystal structures\cite{ziletti_insightful_2018}, but it has the drawback that 2D diffraction fingerprint is not unique across space groups. 
Recently, several papers proposed to use deep learning models trained on simulated CXDPs to perform phase retrieval \cite{Cherukara2018,chan2020realtime,Wu:cw5029,doi:10.1063/5.0014725,wu_3d_2021} which is commonly carried out using iterative algorithms. This demonstrates the emergence of deep learning in the field of CXD and BCDI.

 In this work, we develop and train a 3D convolutional neural network (CNN), which aims to obtain a fast and precise defect classification in nanocrystals of common face-centered cubic (fcc) transition metals. The training data are generated from atomistic simulations that are representative of the physics of the material. Once trained, the network can predict dislocations on simulated and measured 3D  CXDPs. The  predictions are categorized in two (defect free and single dislocation) or three (defect free, single screw and edge dislocations) classes. This work paves the way for automated defect detection and its reliable recognition from 3D CXDPs.

\section*{Results and discussion}
\subsection*{Building the datasets}

\begin{figure}[ht!]
\centering
\includegraphics[width=0.99\textwidth, angle=0]{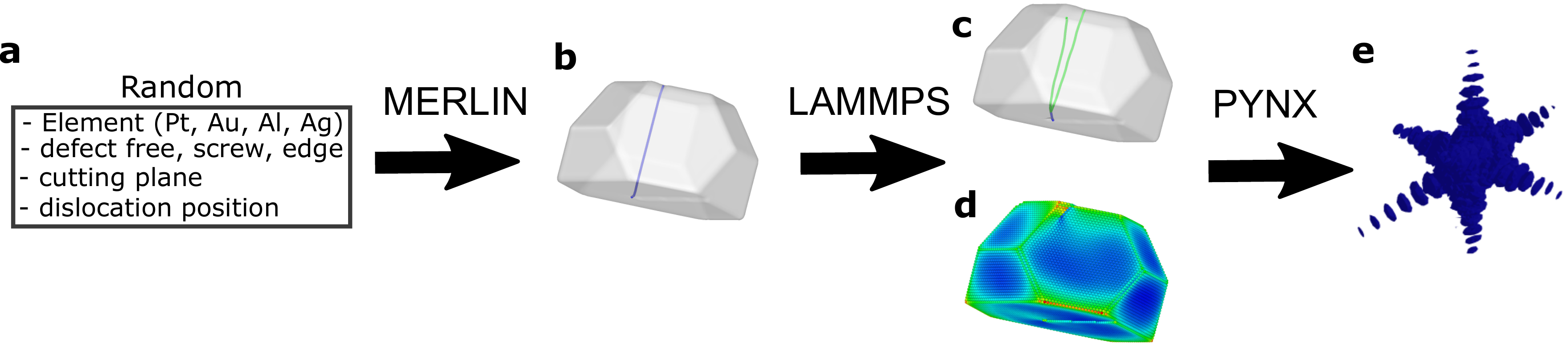}
\caption{\textbf{Schematics of the framework for creating 3D simulated datasets}. \textbf{a} A random element, class, crystal surfaces and dislocation position are selected. \textbf{b} The corresponding crystal is constructed using MERLIN\cite{Rodney2010} and relaxed using LAMMPS\cite{lammps}. \textbf{c} The dislocation dissociates into Shockley partials and \textbf{d} a strain field builds-up in the crystal. \textbf{e} Finally, the corresponding 3D CXDP is calculated with PyNX\cite{Favre-Nicolin2011pynx}.}
\label{fig:pipeline}
\end{figure}

In order to build the dataset required for training the neural network (NN), several material simulation tools were used. The data pipeline allows one to generate simulated CXDPs very close to the ones obtained from Bragg CXD experiments. Fig. \ref{fig:pipeline} illustrates our approach for the creation of 3D CXDPs. The geometry considered in this study is derived from the Wulff construction, \textit{i.e.}, the equilibrium crystal shape of a free-standing crystallite obtained by Gibbs thermodynamic principle, which minimizes the total surface free energy associated to the crystal-medium interface.  \cite{miracle2013wulff} 
In order to take into account the presence of a solid-solid interface, \textit{i.e.} the presence of an underlying substrate as in the experimental nanoparticles, the so-called Winterbottom shape, which can be described as a truncated Wulff construction, is employed.\cite{Winterbottom1967}. An example of a simulated crystal is shown in Fig. \ref{fig:pipeline}b-d. Only fcc transition metals are considered in this study (Al, Au, Ag, Pt), for which the Wulff/ Winterbottom geometries mostly consist of \{1 1 1\} and \{1 0 0\} facets. The Winterbottom constructions are generated using the atomistic simulation code MERLIN \cite{Rodney2010}, by creating a cube of atoms and cutting it along the <1 1 1> and <1 1 0> crystallographic directions, the position of the cut planes being defined by the ratio of the surface energies $\gamma_{111}$ / $\gamma_{100}$ and $\gamma_{110}$ / $\gamma_{100}$ of the material/potential of interest. The lattice orientations corresponding to the axes of the simulation cell are x[1 0 0], y[0 1 0] and z[0 0 1] and are kept constant for all configurations. The interface plane is selected randomly among the eight possible \{1 1 1\} planes, and is cut at a random position corresponding to 65\% - 75\% of the height of a free standing Wulff particle.

Two crystal sizes are considered in this study, the small crystals consist of 40x40x40 unit cells (Supplementary Figure 1) while the large crystals are made up of 80x80x80 unit cells (Supplementary Figure 2). This corresponds to a size of 15x15x(9 -12)nm$^3$  / 100000-140000 atoms for the small configurations, and 30x30x(19-25)nm$^3$  / 800000-950000 atoms for the large configurations, the height and number of atoms in the the crystal depend on the distance of the interface plane with respect to the centre of the particle, and on the lattice parameter of the element considered. For the purpose of this study, we focus on line defects, namely, edge and screw dislocations. A single dislocation and its corresponding displacement field (hypothesis of an isotropic and semi-infinite volume, see Ref. \cite{dupraz2015signature}) is introduced following two strategies. In the first type of configurations, hereafter referred as CD, the dislocation is systematically introduced close to the centre, within a range not exceeding 10\% of the lateral size of the particle. In the second type of configurations, hereafter referred to as RPD, the position of the dislocation is completely random. The simulated dislocations have a Burgers vector of \textbf{b} = $\frac{1}{2}$[1 $\bar{1}$ 0] which is kept constant for all the configurations. This implies that the initial line directions are \textbf{t} = [1 $\bar{1}$ 0] and \textbf{t} = [1 1 $\bar{2}$] for the screw and edge dislocations, respectively. If the Burgers vector and line direction of the dislocations are not varied, the random selection of the interface plane ensures that a large variety of orientations of the dislocation line with respect to the normal of the interface plane is available in the dataset as shown in Supplementary Figures 1 \& 2.

\paragraph{} Once the atomistic configurations are generated, the next step is to obtain accurate and realistic relaxed configurations that reproduce as faithfully as possible the displacement fields measured in the experimental particle. To do so, Molecular Statics simulations are carried out with the open-source Large-scale Atomic/Molecular Massively Parallel Simulator (LAMMPS) \cite{lammps}. The interaction between atoms are modelled with different embedded-atom model (EAM) potentials that accurately reproduce elastic properties as well as surface and stacking fault energies, parameters that are essential to get an accurate description of the relaxed defects. For Al, Ag, Au and Pt we use the EAM potentials developed by Mishin \textit{et al.} \cite{Mishin1999}, Williams \textit{et al.} \cite{Williams2006}, Grochola \textit{et al.} \cite{Grochola2005} and Zhou \textit{et al.} \cite{Zhou2004} respectively. 

The crystals are relaxed at 0 Kelvin using a conjugate gradient algorithm. If the dislocations introduced close to the centre of the nanocrystals are stabilized by the image forces during relaxation, one notable challenge is the tendency of the dislocations introduced close to a free surfaces to escape the crystal during the energy minimization. In order to prevent this phenomenon, the energy tolerance is used as the main stopping criterion for the energy minimization. The latter is defined as the energy change between two successive iterations divided by the total energy of the system, and is set to a value of $10^{-6}$ for the RPD configurations. This value is sufficiently high to ensure that the dislocations dissociate into Shockley partials and remain in the crystal at the end of the relaxation, as shown in Fig. \ref{fig:pipeline}c and Supplementary Figures 1, 2 \& 3. The small number of minimization steps also prevents large rotations of the dislocations during the relaxation. It was indeed observed that edge dislocations are prone to rotate (thus becoming a mixed dislocation) during the energy minimization, especially when they are introduced in the vicinity of the free surfaces. Limiting the number of relaxation steps allows to retain the edge and screw character of the dislocation  during the relaxation, even if dislocations very close to the free surfaces tend to have a mixed character as illustrated in Supplementary Figures 1, 2 \& 3. Each dataset typically contains 1000 relaxed configurations with one third of defect free nanocrystals, one third containing a relaxed screw dislocation and the last third with a relaxed edge dislocation. The time required for the energy minimization of a full dataset ranges between 10 and 25 minutes for the small crystal dataset and 1h30 minutes and 4h for the large crystal dataset.  

\paragraph{} The last step in the dataset creation is the calculation of the three-dimensional CXDPs that are used as input data for our CNN. This is done by summing the amplitudes scattered by each atom with its phase factor, following a kinematic approximation: 
\begin{equation}
\ I(\textbf{q})= \lvert \sum_{j} f_j(q) e^{-2\pi i \textbf{q.r}_j } \rvert^2,
\label{eq:eq1} 
\end{equation}

where \textbf{q} is the scattering vector, $f_{j}$(q) and $r_{j}$ are respectively the atomic scattering factor and position of atom \textit{j}. Note that the crystallographic convention is used in this manuscript, \textit{i.e.} the 2$\pi$ factor is not included in $q$, which implies that a given $q$ value corresponds to a real space distance $d$ of $q = 1/d$. The computation is performed with a GPU using the PyNX\cite{Favre-Nicolin2011pynx} scattering package, which considerably speeds up the calculation of the CXDPs. Given the large number of atoms ($10^5$ – $10^6$ atoms) and the large number of CXDPs that are generated for each dataset (2000-15000), the calculations are performed on 64x64x64 reciprocal space points. The size of the 3D array is a trade-off between achieving a high-enough resolution in the reciprocal space, which is required for an accurate comparison with the experimental CXDPs and keeping the time required to generate the dataset reasonable. Using a POWER9 machine, each CXDP is calculated in ~0.25s for the small configurations and ~2s for the large configuration. A dataset containing 10000 CXDPs is therefore typically generated in 40 minutes for the small nanocrystals and 6 hours for the large ones.
\paragraph{} In order to introduce enough variation in the dataset and prevent overfitting of the model to the training set (Supplementary Figure 4), each CXDP is rotated randomly around the chosen $\vec{Q}$ vector, typically we consider 10 random orientations for each relaxed configuration. The reciprocal space sampling ($\delta$q) is also varied, which is equivalent to zooming around the Bragg reflection of interest (Supplementary Figure 14). A low reciprocal space resolution (coarse sampling / large $\delta$q) can have detrimental effects on the accuracy of the network predictions (Supplementary Figure 14c). To prevent this loss in accuracy, we typically selected $\delta$q values for which the oversampling ratio is consistent with the one used for experimental data. Note that even for the largest $\delta$q values, the oversampling criteria as defined by Sayre \cite{Sayre1952} are still fulfilled, as it is always the case for the experimental CXDPs. Since the simulated particles are significantly smaller than the experimental ones (typically by one order of magnitude), this also implies that a larger portion of the Brillouin zone is selected for the simulated particles. We will see in the following that this has little consequence for the accuracy of the network predictions.
Before training the NN, the distribution of dislocation positions is typically estimated by comparing the maximum of the intensity scattered by the atomistic configurations in the dataset with the maximum of the intensity scattered by a defect free crystal with a similar number of atoms (Supplementary Figure 15). 

\subsection*{Convolutional neural network}

\begin{figure}[ht!]
\centering
\includegraphics[width=0.9\textwidth, angle=0]{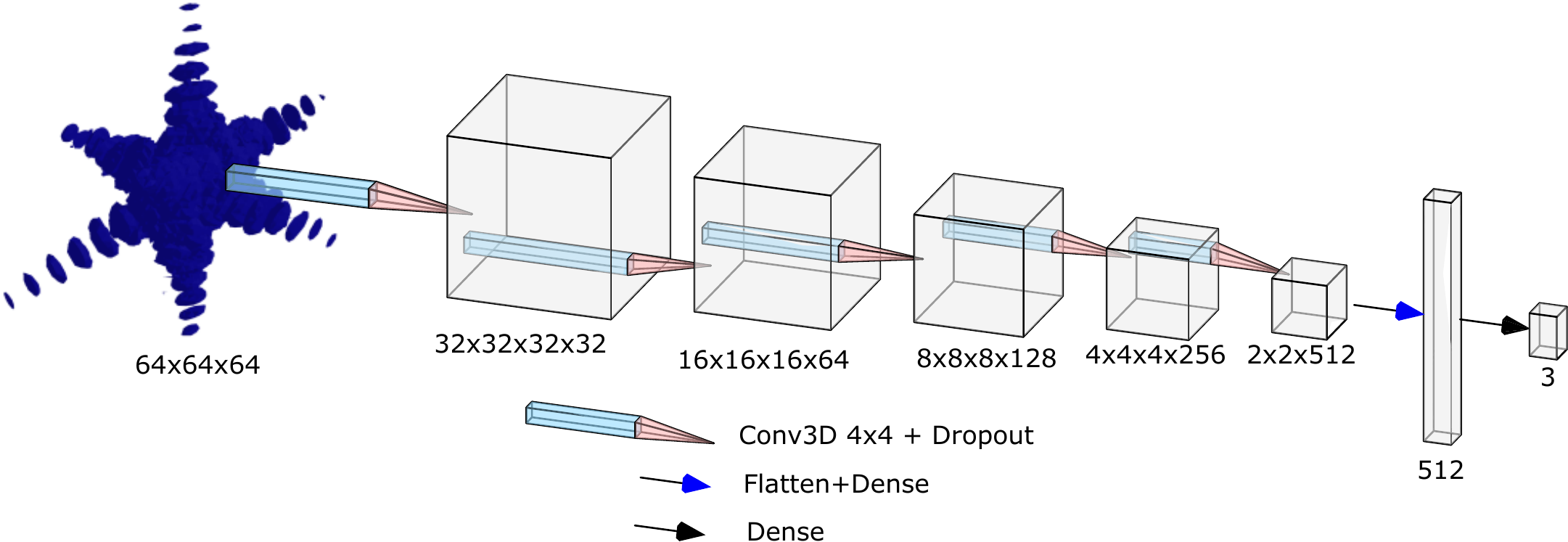}
\caption{\textbf{Schematics of the neural network structure}.The NN model consists of five convolution layers with Relu activation function. We use dropout \cite{dropout} in all layers with a dropout rate of 0.2, to avoid overfitting. 
The model ends with 2 fully connected layers with a last softmax activation function.The model takes the 3D 64x64x64 diffraction as input and predicts the probability for each class (defect free, screw, edge).}
\label{fig:NN}
\end{figure}

The NN model architecture is displayed in Fig. \ref{fig:NN}. It takes as input the 64$\times$64$\times$64 image of the CXDP intensity and encodes it through a series of convolution and fully connected layers. Dropout \cite{dropout} is used in all layers with a dropout rate of 0.2, to avoid overfitting. This is a standard architecture, nevertheless it already gives very accurate predictions on the simulated dataset. Increasing the size of the model, adding extra layers or increasing the number of filters in the convolution layers  does not increase the model efficiency and even leads to an overfit of the training dataset in some cases.

Training is performed using Adam optimization\cite{kingma2017adam} with a learning rate of 10$^{-3}$ and a batch size of 64. A large amount of 3D datasets are simulated. They systematically specify the correct output (defect class) for a given input (3D CXDP intensity), and minimise a categorical cross-entropy loss that quantifies the difference between the predicted and the correct class labels (defect free, screw and edge). Through this minimisation, the weights (\textit{i.e.}, parameters) of the neural network are optimised to reduce the classification error. The weights of each convolutional and fully connected layers are initialized randomly. Moreover, the instances of the training dataset are processed in a random order. Nonetheless, two independent trainings for a given dataset a CNN always gives a very similar probability distribution as illustrated in Supplementary Figure 16.
The simulated data are split into training, validation and test sets. The model fit is performed with the training set and stopped when the validation set accuracy reaches a maximum. The final model prediction on the test set containing 11556 CXDPs calculated from 1284 atomistic configurations reaches a very high total accuracy score of 97.2\%. In addition, the confusion matrix displayed in Supplementary Figure 7 shows that almost all defect free crystals are  predicted. Most of the errors (4.7\%) come from edge dislocations predicted as screw. Furthermore, as illustrated in Supplementary Figures 11 \& 12, a simpler two classes model (Supplementary Figure 10) predicting either a defect free oe a defective crystal can reach an even higher accuracy. From an occlusion sensitivity test\cite{zeiler2014visualizing} on a simulated CXDP shown in Supplementary Figure 13, we demonstrate that the NN mainly uses the vicinity of the Bragg peak to make its prediction. 


\subsection*{Validation on experimental data}

\begin{figure}[ht!]
\centering
\includegraphics[width=0.8\textwidth, angle=0]{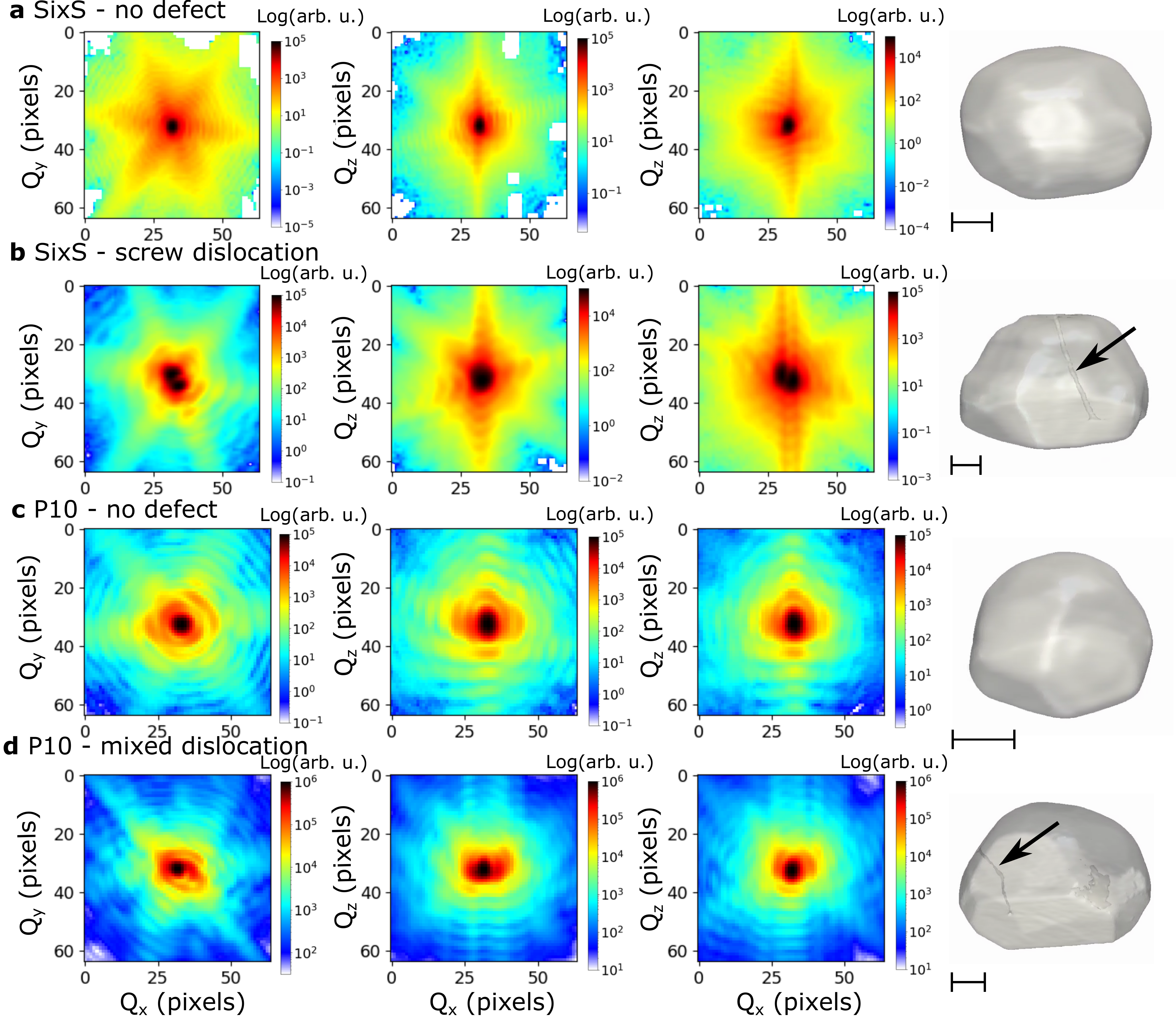}
\caption{\textbf{CXDPs and reconstructed Bragg electron density from Pt NPs measured on the P10 and SixS beamlines.} \textbf{a} Defect free NP, SixS beamline \textbf{b} Defective NP, SixS beamline \textbf{c} Defect free NP, P10 beamline \textbf{d} Defective NP, P10 beamline. The black arrows indicate the location of the dislocations. The scale bar indicates a length of 100 nm. The CXDPs are on a logarithmic scale to enhance the fringes visibility}
\label{fig:experience}
\end{figure}

The experimental datasets correspond to 3D reciprocal space maps obtained by measuring the Bragg CXDPs of Pt nanoparticles. Single particles were measured either at the SixS beamline of synchrotron SOLEIL (Orsay, France) or at the P10 beamline of synchrotron PETRA (Hamburg, Germany). The 3D Bragg CXDPs were collected at the asymmetrical $\bf{\bar{1}11}$ Pt Bragg reflection at the SixS beamline or at the symmetrical (specular) \textbf{111} Pt Bragg reflection at the P10 beamline. The experimental reciprocal space datatsets have been orthonormalised using the xrayutilities package \cite{kriegner_xrayutilities_2013}. Fig. \ref{fig:experience} displays the CXDPs of the experimental datasets, as well as their reconstructed Bragg electron density using phase retrieval algorithms. Defect-free (Figs. \ref{fig:experience}a,c) as well as defective crystals (Figs. \ref{fig:experience}b,d) were measured. A closer look at Figs. \ref{fig:experience}(b,d) reveals the variety of dislocation configurations that is found in experimental nanocrystals. These dislocations were most likely nucleated during the growth of the nanoparticles, and did not escape during the annealing at 1100$^\circ$C, suggesting that they are strongly pinned in the nanocrystal. For the SixS data, the screw dislocation is close to the center of the nanocrystal (Burgers vector of \textbf{b} = $\frac{1}{2}$[1$\bar{1}$0]). On the other hand, the dislocation in the P10 defective nanocrystal is closer to the free surfaces. In addition, the dislocation line is not perfectly straight and parallel to the Burgers vectors (\textbf{b} = $\frac{1}{2}$[101]). It can thus be described as a mixed dislocation  with a dominant screw character. 

In order to reinforce the agreement between the simulated and experimental datasets each diffraction measurement is preprocessed before computing the model prediction. The CXDP center of mass is placed at the center of the array, as it is also the case for the simulated data. Finally, the CXDP is normalized so that the maximum is equal to 1.

\begin{figure}[ht!]
\centering
\includegraphics[width=.75\textwidth, angle=0]{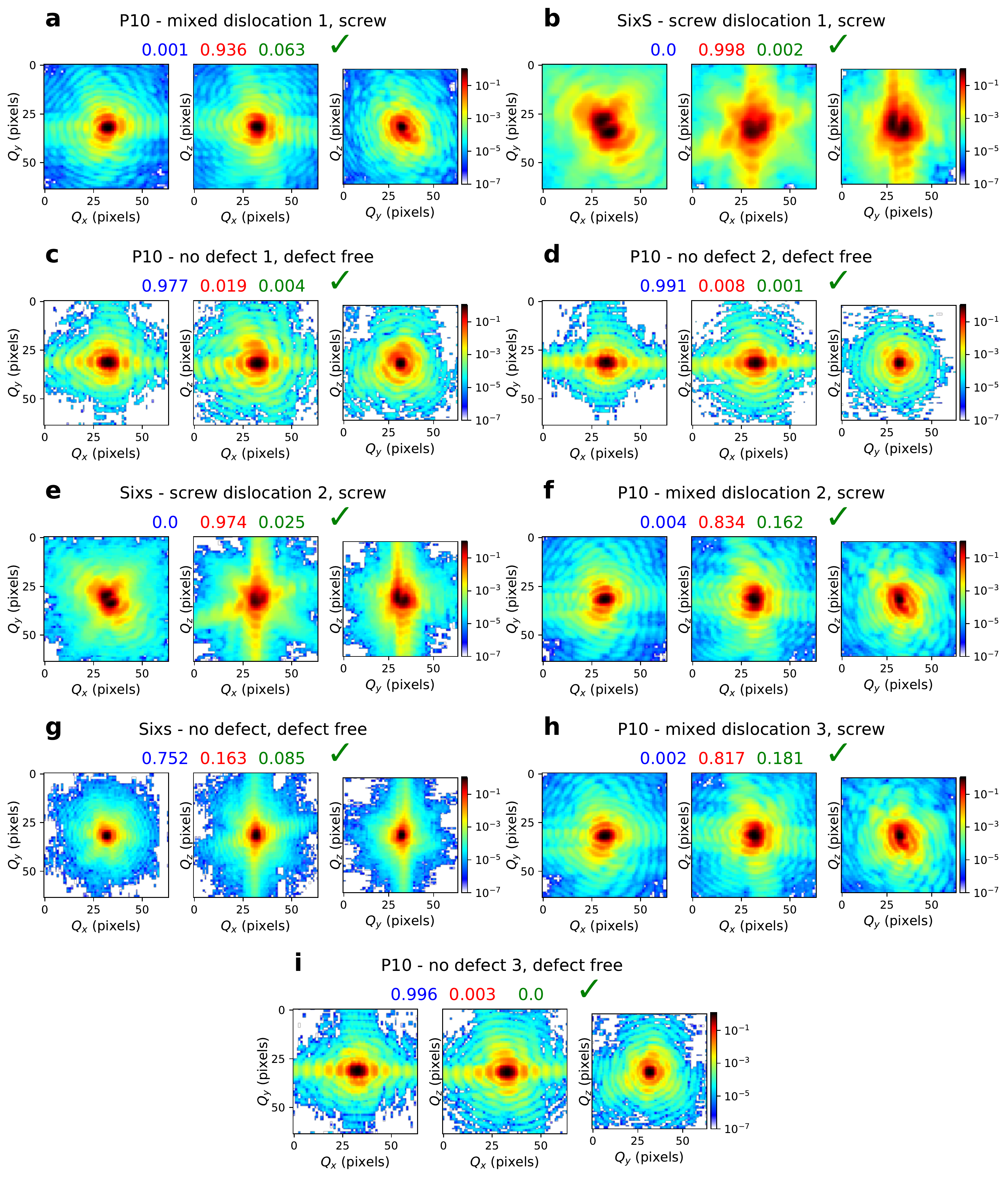}
\caption{\textbf{Predictions from the best model on experimental data}. For each example, the probabilities for the 3 classes are shown in the title where blue, red and green correspond respectively to defect free, screw and edge class. Three cross-sections of the 3D CXDP are shown for each example. Using this NN model, all examples are correctly predicted (see green ticks). The CXDPs are on a logarithmic scale to enhance the fringes visibility}
\label{fig:results}
\end{figure}

The results of our best NN model on the preprocessed CXDPs are displayed in Fig. \ref{fig:results} along with slices along $Q_x$, $Q_y$ and $Q_z$ for each experimental CXDP. Some crystals were measured several times  under different experimental conditions (temperature, gas environment) for example P10 - no defect 1, 2 and 3 in Fig. \ref{fig:results}), allowing us to compare the model predictions for the same crystal but with slightly different CXDPs.

The performances of this model on experimental data are excellent, all the experimental examples being predicted in the correct class and most of them with a very high probability (> 95\%). Although still very good, the predictions for the P10 data (mixed dislocation) are generally slightly worse with an accuracy ranging between 82 and 94\%. This is not surprising given the mixed type of dislocation (with a dominant screw character), which necessarily increases the probability of identifying the defect as an edge dislocation. Nonetheless, even if the dislocation is located close to a free surface and therefore induces weak distortions in the CXDP (Fig. \ref{fig:experience}d), our model still manages to identify this crystal as defective with almost a 100\% probability. This demonstrates the robustness of the model trained on this dataset, which can predict both centered and off-centered dislocations with a very high accuracy. 

The simulated training dataset used to fit the NN model has a large influence on the accuracy of the predictions on experimental data. This dataset must contain enough diversity, while sharing enough similarities with the experimental CXDPs. The predicted probabilities on experimental data for the same model architecture but different simulated training datasets are shown in Table \ref{table:predictions}. Six different simulated datasets have been trained: (1) single element (Pt) unrelaxed small crystals, 100\% centered dislocations (CD), (2) relaxed Pt small crystals (100\% CD), (3) relaxed Pt large crystals (100\% CD), (4) relaxed large crystals with multiple elements (Au and Pt) (100\% CD), (5) relaxed multi-elements large crystals with dislocations at random position (100\% RPD) and (6) relaxed multi-elements large crystals with a mix of CD and RPD configurations (75\% CD and 25\% RPD). The first two rows of Table \ref{table:predictions} emphasize the importance of accurately modelling the displacement field of the dislocations. Indeed, while these two models trained on relaxed and unrelaxed datasets predict accurately the defect free configurations, they fail at identifying the mixed dislocation (P10 data). However, the model trained on the relaxed dataset performs much better on the SixS-"screw" data, which is correctly identified as a screw dislocation (see also Supplementary Figure 9). On the other hand, the size of the relaxed crystals does not have a major impact on the accuracy of the model (Table \ref{table:predictions}, second and third row), although the predictions of the models trained on the large configurations is slightly more accurate, in particular for defect free configurations (Supplementary Tables 2 and 6). 
\paragraph{} The addition of several elements in the dataset improves the accuracy of the predictions for the SixS data, but has no effect on the P10 data (Table \ref{table:predictions}, fourth row). Nonetheless, mixing several elements in the dataset generally results in better model predictions compared to the models based on single elements, in particular for the large crystal size (Supplementary Tables 5-8). The position of the dislocation also has a major impact on the model predictions. As seen from Table \ref{table:predictions} (sixth row), introducing the dislocation at random positions, including positions close to the crystal free surfaces, results in more accurate predictions for the P10 data. However, this improvement is at the expense of the predictions for the SixS data, which is correctly identified as a dislocation, but with an edge character instead of a screw. The predictions for the defect free data are not affected and still excellent (see also Supplementary Figure 8). 
\paragraph{} In order to obtain accurate predictions simultaneously for both P10-"mixed" and SixS-"screw" dislocations, one must increase the diversity in the training dataset. This has been achieved by building a dataset consisting of a mix of CD and RPD configurations (Table \ref{table:predictions}, seventh row). Training the CNN on this mixed dataset significantly enhances the performances of the model and allows to predict correctly and with a very high accuracy all the experimental examples.

We must emphasize that, despite the differences in the ability of the models to generalize to experimental data, the accuracy on the simulated test data for each training dataset is always higher than 86\% (Supplementary Tables 1 \& 5). Our work illustrates the necessity of using a simulated trained datasets close to real structures: atomistically relaxed nanoparticles with an accurate modelling of the dislocation displacement field, multiple atomic elements and random location of the dislocations. It also demonstrates that a convolutional neural network can predict dislocations in a crystal from its 3D coherent diffraction pattern. Combined with the fast scanning capabilities of some synchrotron beamlines \cite{Chahine2014}, this approach could be used to perform a fast screening of the nanocrystals on a sample of interest. This would allow to determine the proportion of defect free nanocrystals as well as nanocrystals containing a specific type of crystal defect, and select the best candidate for a coherent diffraction imaging experiment. In addition, if the CNN was only tested on metallic fcc particles, we foresee that it could be extended to more complex systems like for instance multi-element particles.

\paragraph{} From 3D coherent X-ray CXDPs, we used a convolutional neural network to predict defect classes. As a result, we obtain an automatic procedure for defect classification in fcc metals, which does not require any user-manipulation, any intensive live data mining, and achieves high-accuracy classification even in the presence of defects close to the free surfaces of the nanocrystals. This tool can be exploited during experiment execution to provide rapid feedback to the investigator, enables one to identify on the fly target defect types present in individual nanocrystals, and furthers the possibility of unsupervised data collection, extremely relevant given the increases data rates expected at ever improving facilities. Our study paves the way for defect recognition of three-dimensional structural data in big-data materials science.

\newpage
\section*{Methods}
\subsection*{Training the network}

We used the python deep-learning API Keras \cite{chollet2015keras} running the TensorFlow backend \cite{tensorflow2015-whitepaper} to build, develop and train our NN. The training was performed in parallel on two NVIDIA Tesla V100 GPUs and a POWER9 computer. 
We use a categorical cross-entropy loss function $L(y,\hat{y}) = -\frac{1}{B}\sum\limits_{n=1}^{B}\sum\limits_{c=1}^{N_c}y_{n,c}\,\text{log}(\hat{y}_{n,c})$ where $B$ is the batch size, $N_c$ the number of classes, $y_{n,c}=1$ for data element $n$ if the true class is $c$ and $y_{n,c}=0$ otherwise. $\hat{y}_{n,c}$ is the predicted probability for class $c$. The simulated dataset is divided into training, validation and test, corresponding respectively to 85\%, 10\% and 5\% of the total dataset. The model is trained with a learning rate of $10^{-3}$ and a batch size of 64 on the training set until the model accuracy calculated on the validation set reaches a plateau (Supplementary Figures 5, 6 \& 11). A typical training requires between 15 and 30 minutes depending on the dataset (8-10 seconds per epoch and 100-200 epochs). Decreasing the learning-rate and increasing the batch size does not further improve the model accuracy. Once trained, the model performance is evaluated on the test set and reaches a total accuracy >86\% on the simulated data for all models presented in Table \ref{table:predictions}.

\subsection*{Sample growth}
Pt nanocrystals were prepared by the solid-state dewetting of a 30-nm thin Pt film for 24 hours at 1100$^\circ$C in air\cite{li_continuous_2020}. The Pt film was deposited on $\alpha$-Al$_2$O$_3$ (sapphire) with an electron beam evaporator. The Pt nanocrystals have their \textit{c}-axis oriented along the [111] direction normal to the (0001) sapphire substrate. A standard photolithography method was employed to prepare a patterned layer of photoresist on sapphire prior to the electron beam evaporation of Pt. The lithographic processing route ensured that a number of dewetted Pt particles are well-separated from their neighbors and that only one crystallite is irradiated by the incoming x-ray beam. The particle size ranges from 100 nm to 700 nm.

\section*{Data availability}

The data supporting the findings of this work are available from the corresponding author on reasonable request.

\section*{Acknowledgements}

We acknowledge financial support from the European Research Council (ERC) under the European Union’s Horizon 2020 research and innovation programme (grant agreement No. 818823). We also thank the support by a grant from the Ministry of Science \& Technology, Israel and CNRS, France.

\section*{Competing interest}

The authors declare no competing interests.

\section*{Author contributions statement}

B.L. and M.D. performed the atomistic simulations. E.B. developed the convolutional neural network. M.-I.R. processed the experimental data. E.R. and E.A. prepared the samples. M.D., M.-I.R, S.L, A.R., A.C. performed the experiments at the SixS beamline of synchrotron SOLEIL (experiment number: 20181680) and M.D., M.-I.R, S.L and M.S. performed the experiments at the P10 beamline of synchrotron PETRA (experiment number: I-20180962). B.L., E.B., M.D. and M.-I.R. wrote the manuscript and all authors reviewed the manuscript. B.L., E.B. and M.D. contributed equally to this work and are considered "co-first author" of this manuscript.

\section*{Additional information}

Supplementary Information is available for this paper at XX.

\newpage



\input{ms.bbl}
\section*{Figures and Tables}

\begin{table}[ht!]
\begin{tabular}{|c|c|c|c|c|c|c|c|c|c|}\hline
\diaghead{\theadfont Diag ColumnmnHead II}%
{Training\\dataset}{Experimental\\example}&
\thead{P10\\mixed 1}&\thead{SixS\\Screw 1}& \thead{P10\\no defect 1}& \thead{P10\\no defect 2}& \thead{SixS\\Screw 2}& \thead{P10\\mixed 2}& \thead{SixS\\no defect}& \thead{P10\\mixed 3}& \thead{P10\\no defect 3}\\

\hline
\thead{Pt\\unrelaxed small crystals \\ CD }&   \redn{\thead{p 99\\s 1\\ e 0}}&    \redn{\thead{p 0\\s 3\\ e 97}}&    \grn{\thead{p 100\\s 0\\ e 0}}&    \grn{\thead{p 100\\s 0\\ e 0}}&    \redn{\thead{p 0\\s 1\\ e 99}}&    \redn{\thead{p 97\\s 2\\ e 1}}&    \grn{\thead{p 100\\s 0\\ e 0}}&    \redn{\thead{p 100\\s 0\\ e 0}}&    \grn{\thead{p 100\\s 0\\ e 0}}\\

\hline
\thead{Pt\\relaxed small crystals \\ CD }&   \redn{\thead{p 99\\s 1\\ e 0}}&    \grn{\thead{p 0\\s 100\\ e 0}}&    \grn{\thead{p 100\\s 0\\ e 0}}&    \grn{\thead{p 100\\s 0\\ e 0}}&    \grn{\thead{p 0\\s 69\\ e 31}}&    \redn{\thead{p 72\\s 18\\ e 10}}&    \grn{\thead{p 72\\s 16\\ e 12}}&   \redn{\thead{p 92\\s 6\\ e 2}}&    \grn{\thead{p 100\\s 0\\ e 0}}\\

\hline
\thead{Pt\\relaxed big crystals \\ CD }&\redn{\thead{p 99\\s 0\\ e 1}}& \grn{\thead{p 0\\s 100\\ e 0}}& \grn{\thead{p 100\\s 0\\ e 0}}& \grn{\thead{p 100\\s 0\\ e 0}}& \grn{\thead{p 25\\s 38\\ e 37}}& \redn{\thead{p 99\\s 0\\ e 1}}& \grn{\thead{p 99\\s 0\\ e 1}}& \redn{\thead{p 99\\s 0\\ e 1}}& \grn{\thead{p 100\\s 0\\ e 0}}\\


\hline
\thead{Multi-elements\\relaxed big crystals\\ CD}& \redn{\thead{p 98\\s 0\\ e 2}}& \grn{\thead{p 0\\s 100\\ e 0}}& \grn{\thead{p 100\\s 0\\ e 0}}& \grn{\thead{p 100\\s 0\\ e 0}}& \grn{\thead{p 0\\s 98\\ e 2}}& \redn{\thead{p 88\\s 5\\ e 8}}& \grn{\thead{p 99\\s 0\\ e 1}}& \redn{\thead{p 94\\s 2\\ e 3}}& \grn{\thead{p 100\\s 0\\ e 0}}\\

\hline
\thead{Multi-elements\\relaxed big crystals\\RPD}& \grn{\thead{p 9\\s 53\\ e 38}}& \redn{\thead{p 0\\s 36\\ e 64}}& \grn{\thead{p 100\\s 0\\ e 0}}& \grn{\thead{p 100\\s 0\\ e 0}}& \redn{\thead{p 0\\s 8\\ e 92}}& \grn{\thead{p 0\\s 58\\ e 42}}& \grn{\thead{p 99\\s 1\\ e 0}}& \grn{\thead{p 0\\s 63\\ e 37}}& \grn{\thead{p 100\\s 0\\ e 0}}\\

\hline
\thead{Multi-elements\\relaxed big crystals\\75\% CD, 25\% RPD}& \grn{\thead{p 0\\s 94\\ e 6}}& \grn{\thead{p 0\\s 100\\ e 0}}& \grn{\thead{p 98\\s 2\\ e 0}}& \grn{\thead{p 99\\s 1\\ e 0}}& \grn{\thead{p 0\\s 97\\ e 3}}& \grn{\thead{p 0\\s 84\\ e 16}}& \grn{\thead{p 75\\s 16\\ e 9}}& \grn{\thead{p 0\\s 82\\ e 18}}& \grn{\thead{p 100\\s 0\\ e 0}}\\
\hline
\end{tabular}
\caption{\textbf{Predicted probabilities on the experimental data from several models trained with different simulated training datasets}. The predictions are shown in \%. In each cell, the prediction probability for the 3 classes (prefect: p, screw: s, edge: e) is shown in green if the prediction is correct and in red if it is wrong. CD and RPD stand for centered dislocations and random position dislocations, respectively.}
\label{table:predictions}
\end{table}

\end{document}


\title{Supplementary Information: A Convolutional Neural Network for Defect Classification in Bragg Coherent Diffraction}

\date{\today}

\maketitle


\section{Supplementary Methods}

\subsection{Simulated crystal dataset}

Since the number of atoms in simulated crystal configurations are limited by the computing capabilities of our computers, the simulated crystal size is smaller (diameter < 30 nm) than the one measured experimentally (diameter between 300 nm and 600 nm). 
To ensure that the size of the simulated crystals does not influence the results of the deep learning model, two crystal sizes were considered in this study, hereafter referred as the large and small crystal datasets.

\begin{figure}[ht!]
\centering
\includegraphics[scale = 0.35]{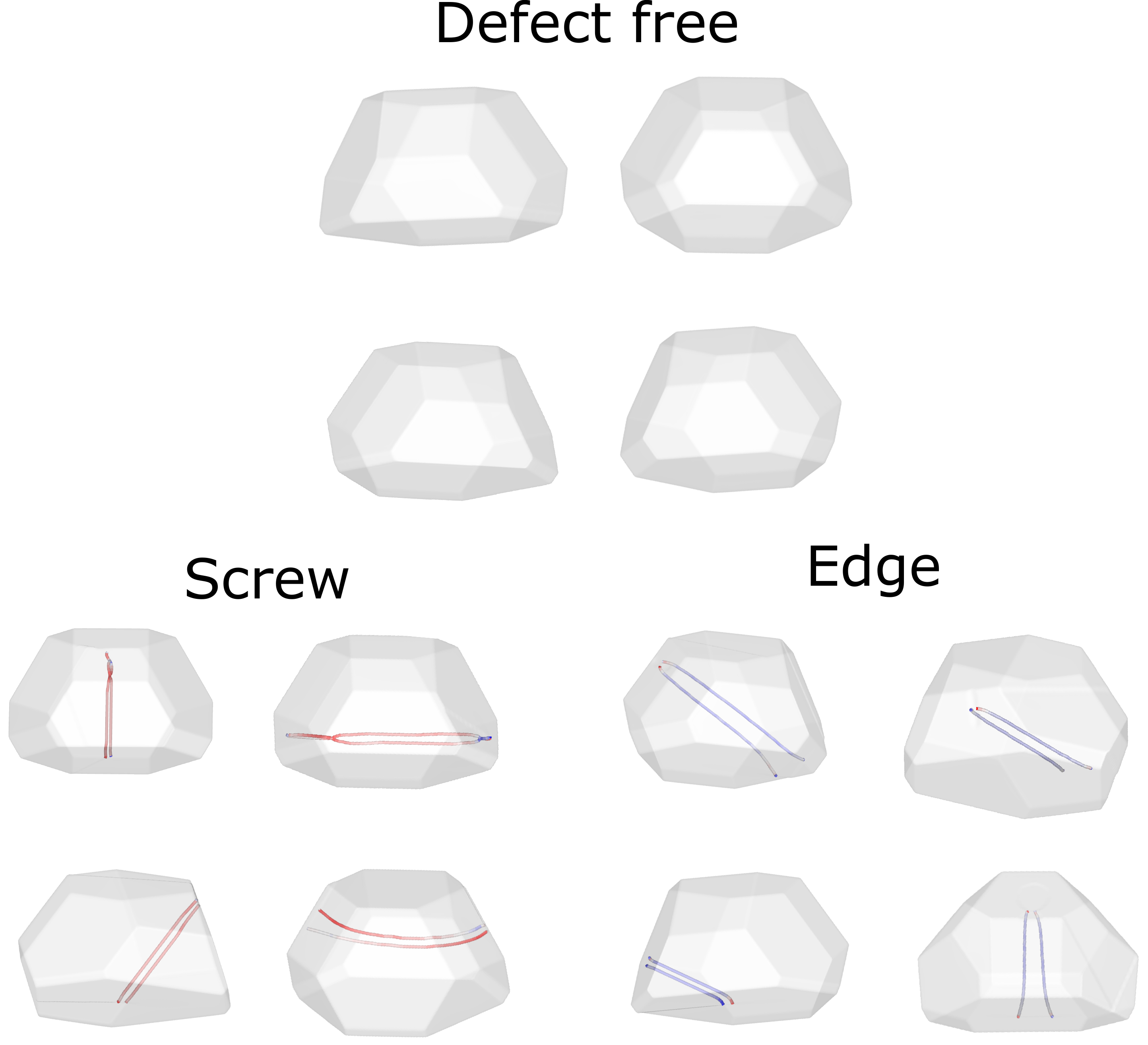}
\caption{\textbf{Atomistic configurations from the large crystal dataset}. They consist of 80$\times$80$\times$80 units cell corresponding to a size of 30$\times$30$\times$(19-25)nm$^3$ / 800000-950000 atoms. The color indicates the local character of the dislocation (blue$=$edge, red$=$screw)}
\label{fig:simulated_examples_large}
\end{figure}

\begin{figure}[ht!]
\centering
\includegraphics[scale = 0.35]{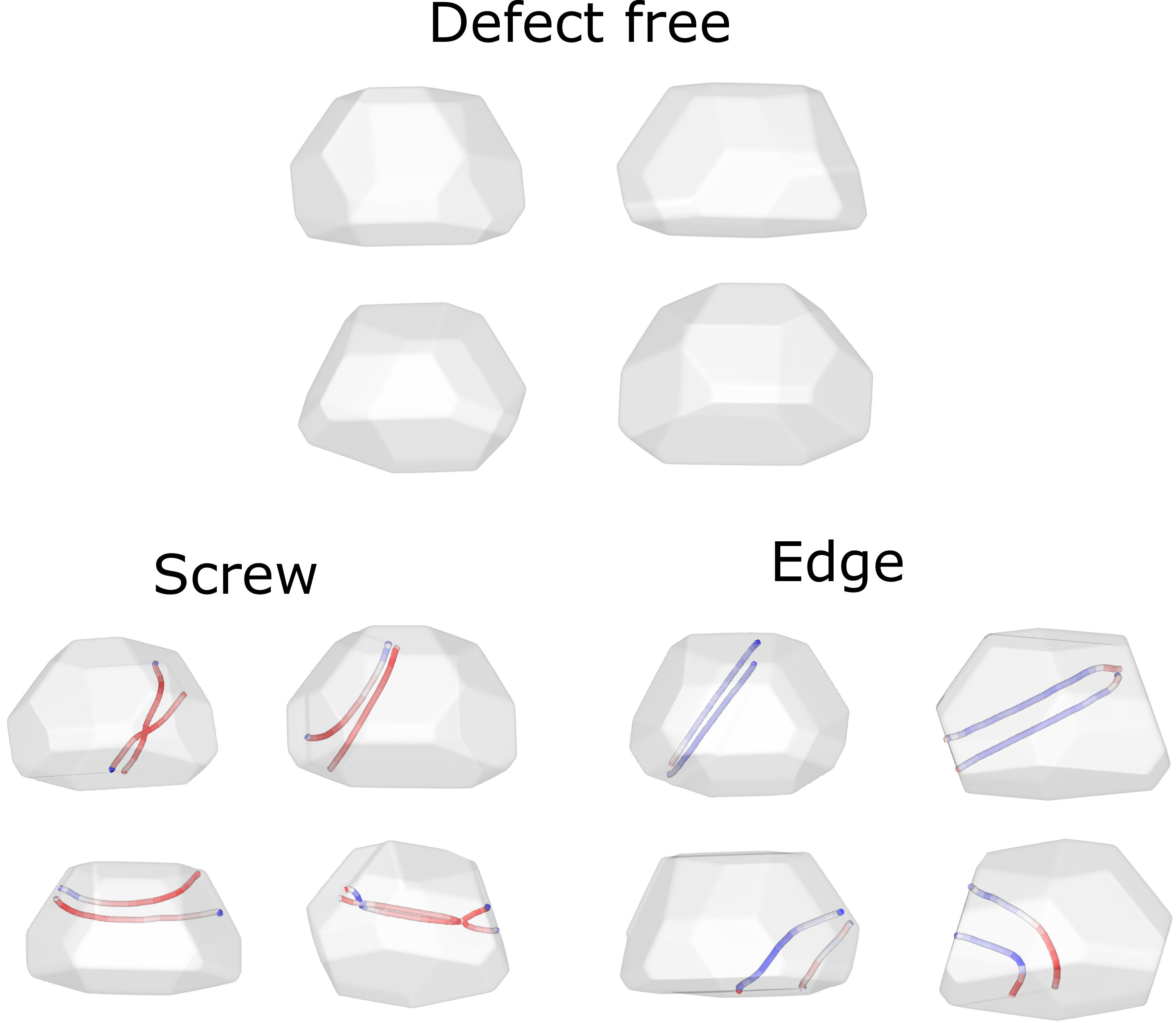}
\caption{\textbf{Atomistic configurations from the small crystal dataset}. They consist of 40$\times$40$\times$40 units cell corresponding to a size of 15$\times$15$\times$(9-12)nm$^3$ / 100000-140000 atoms. The color indicates the local character of the dislocation (blue$=$edge, red$=$screw)}
\label{fig:simulated_examples_small}
\end{figure}

Random examples of relaxed configurations from the large and small crystal dataset in each three classes are shown in Supplementary Figure \ref{fig:simulated_examples_large} and Supplementary Figure \ref{fig:simulated_examples_small} respectively. Perfect dislocations with $\vec{b}$ = $\frac{1}{2}$[1$\bar{1}$0] dissociate into Shockley partials upon relaxation in order to minimize their energy according to Frank's rule \cite{Hull2011}. For edge dislocations, the dissociation is planar and constrained to the (1 1 1) slip plane of the dislocation. The perfect edge dislocation therefore splits into two Shockley partials $\vec{b_{p1}}$ = $\frac{1}{6}$[1$\bar{2}$1] and $\vec{b_{p2}}$ = $\frac{1}{6}$[2$\bar{1}\bar{1}$]. For screw dislocations on the other hand, dissociation can occur in the two available (1 1 1) and (1 1 $\bar{1}$) slip planes, giving rise to one or two sets of partials in the relaxed configurations: $\frac{1}{6}$[1$\bar{2}$1] and $\frac{1}{6}$[2$\bar{1}\bar{1}$] in the (1 1 1) plane and $\frac{1}{6}$[1$\bar{2}\bar{1}$] and $\frac{1}{6}$[2$\bar{1}$1] in the (1 1 $\bar{1}$) plane. As shown in Supplementary Figure \ref{fig:simulated_examples_small}, screw dislocations introduced close to the centre are prone to dissociate into two sets of Shockley partial, forming a constriction node in the process. Because of the negative energy of the constriction node, this crossed configuration has been shown to be more favourable energetically \cite{Rasmussen1997}. The local character of the dislocation is illustrated by a color gradient where blue/red correspond to a local edge/screw character. Since the dislocation curvature is more pronounced, when located close to the crystal surfaces, the dislocation character is generally not as well defined in the small crystal, especially for the crystal at the lower right of Supplementary Figure \ref{fig:simulated_examples_small}.

\begin{figure}[ht!]
\centering
\includegraphics[scale = 0.5]{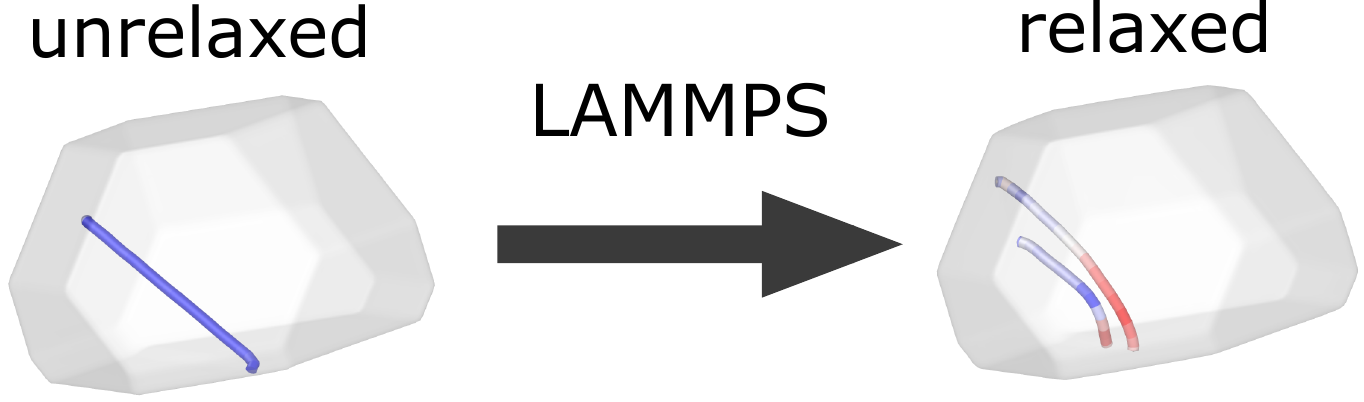}
\caption{\textbf{Relaxation of a small crystal configuration with an edge dislocation}. The color indicates the local character of the dislocation (blue$\rightarrow$edge, red$\rightarrow$screw)}
\label{fig:edge_relax}
\end{figure}

An example of a small simulated crystal with an edge dislocation introduced in the vicinity of the surface is shown in Supplementary Figure \ref{fig:edge_relax}. When this configuration is relaxed using LAMMPS, the dislocation has a mixed character, partially edge in blue and screw in red. The identification of the mixed-dislocation, without any dominant screw or edge character is more difficult. Nonetheless, these defects are often observed in crystals measured experimentally as shown in Supplementary Figure 3d ("P10 - mixed dislocations") and need to be included in the training dataset in order to achieve accurate predictions for these dislocation configurations. Note that in the training dataset, a mixed dislocation is classified as an edge or screw dislocation if the edge or screw segment is dominant, respectively.

\subsection{Model training, augmented dataset and dropout}

\begin{figure}[ht!]
\centering
\includegraphics[scale = 0.7]{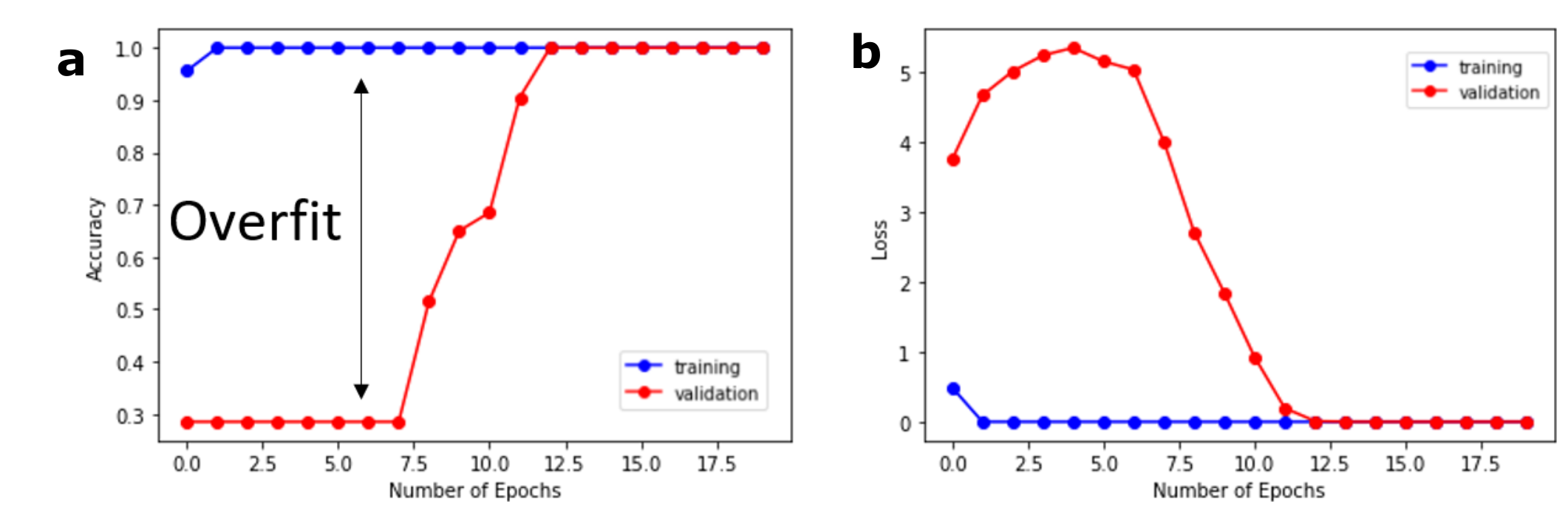}
\caption{\textbf{Illustration of the overfit of the model to the training set}. \textbf{a} Training and validation accuracy of the network with overfit. \textbf{b} Training and validation loss of the network with overfit.}
\label{fig:overfit}
\end{figure}

When training the model with the simulated relaxed crystal configurations, we observed a large difference between the accuracy of the validation set and the one of the training set as shown in Supplementary Figure \ref{fig:overfit}a-b. This is due to an overfit of the model to the training set, a recurrent issue in machine learning. To solve this problem, we introduced more variance in the simulated dataset, including random rotations, random pixel resolution and random subpixel shift of the diffraction pattern.

Training again the model with the augmented dataset results in a better agreement between the training and validation accuracy, and prevents overfitting as shown in Supplementary Figure \ref{fig:no_overfit}a.

\begin{figure}[ht!]
\centering
\includegraphics[scale = 0.7]{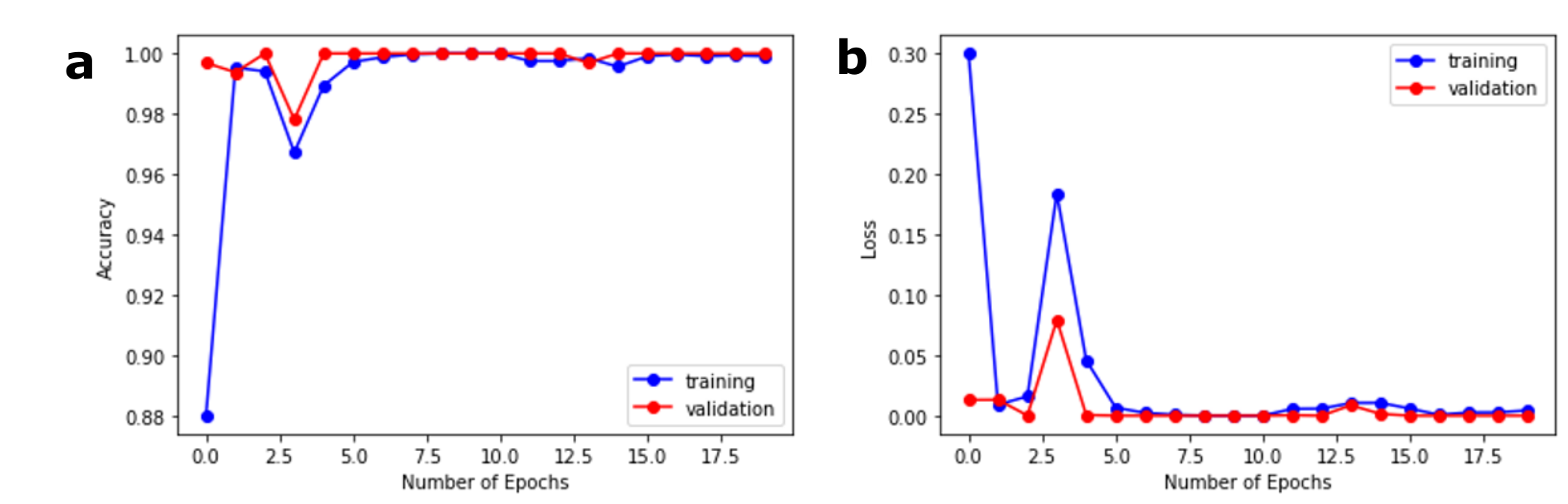}
\caption{\textbf{Training of the NN with the augmented dataset}. \textbf{a} Training and validation accuracy of the network without overfit. \textbf{b} Training and validation loss of the network without overfit.}
\label{fig:no_overfit}
\end{figure}

\begin{figure}[ht!]
\centering
\includegraphics[scale = 0.5]{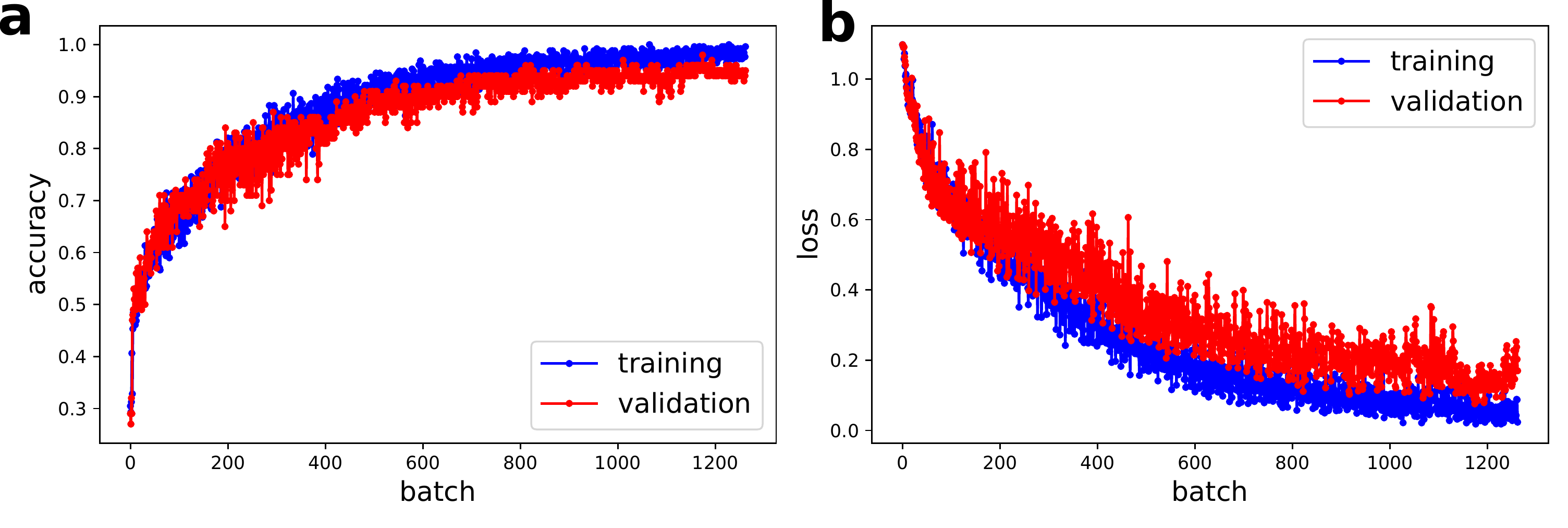}
\caption{\textbf{Detailed evolution of the accuracy and loss function during a training performed on a typical dataset}. \textbf{a} Training and validation accuracy calculated every 256 training examples (batch) instead of a full epoch.  \textbf{b} Same with the training and validation loss}
\label{fig:training_loss_detail}
\end{figure}

Supplementary Figure \ref{fig:training_loss_detail}a and b show a more detailed evolution, during a typical model training, of the accuracy and loss for the validation and training datasets. In this example, both accuracy and loss are calculated every 256 training examples (batch) instead of every epoch. This training was performed with a batch-size of 64 and a learning rate of $10{^-3}$ that was decreased to $5\times10^{-4}$ at the end of the training to stabilize the model. Despite a small difference between training and validation accuracy, overfitting is not observed for this model since the validation accuracy continues increasing until the end of the training.

We added dropout to all layers with a dropout rate of 0.2. This dropout was required to train the model with the unrelaxed dataset as “Pt unrelaxed small crystals CD” in Supplementary Table 1 of the main text. However, dropout is not mandatory for the relaxed dataset, since manually stopping the training when the validation accuracy reaches a maximum gives an equivalent  final score. To confirm this trend, we trained 2 models with a relaxed dataset: first one with dropout used in all layers at a dropout rate of 0.2 and the second one without any dropout. An accuracy score on the same test set reached 95,3\% for the first model and 95,7\% for the second model. Since this difference of 0.5\% is not significant and probably due to randomness during training, we decided to keep dropout on all layers for both unrelaxed and relaxed training datasets.


\newpage 

\section{Supplementary discussion}

\subsection{Confusion matrix for the 3 classes model}

\begin{figure}[ht!]
\centering
\includegraphics[scale = 0.7]{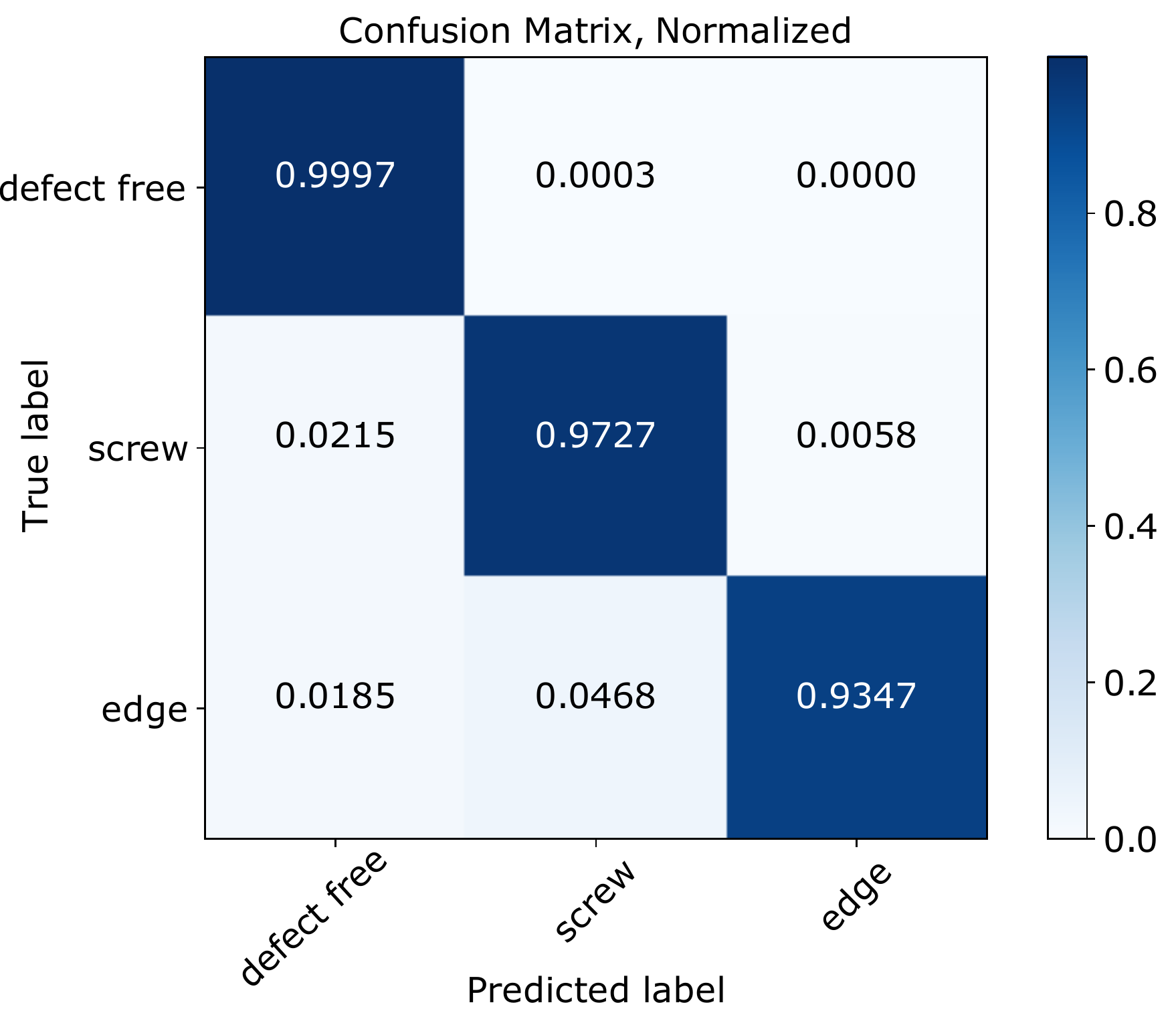}
\caption{\textbf{Normalized confusion matrix for the best 3 classes model presented in the main text}. This matrix was calculated with the simulated test dataset.}
\label{fig:3classes_confusion}
\end{figure}

The normalized confusion matrix for the best 3 classes model shown in the main manuscript is displayed in Supplementary Figure \ref{fig:3classes_confusion}. The total accuracy of this model on the test dataset is 97.2\% but the confusion matrix reveals more information. One can observe that all the defect free crystal examples are predicted correctly while the error is larger for the edge and screw dislocations. Furthermore, the edge examples with a wrong prediction are mostly predicted as screw and not as perfect crystals. This again shows that, while our 3 classes model has a high prediction accuracy, it is even better at making the distinction between a defect free and defective crystal.


\subsection{Model results for different simulated training datasets}

The performances of the CNN on experimental data strongly depend on the simulated training dataset used to fit the model. A good training dataset should contain configurations close to the experimental data but also enough diversity to be able to predict correctly a wide range of experimental cases that can differ significantly from each other.

\begin{figure}[ht!]
\centering
\includegraphics[width=.75\textwidth,angle=0]{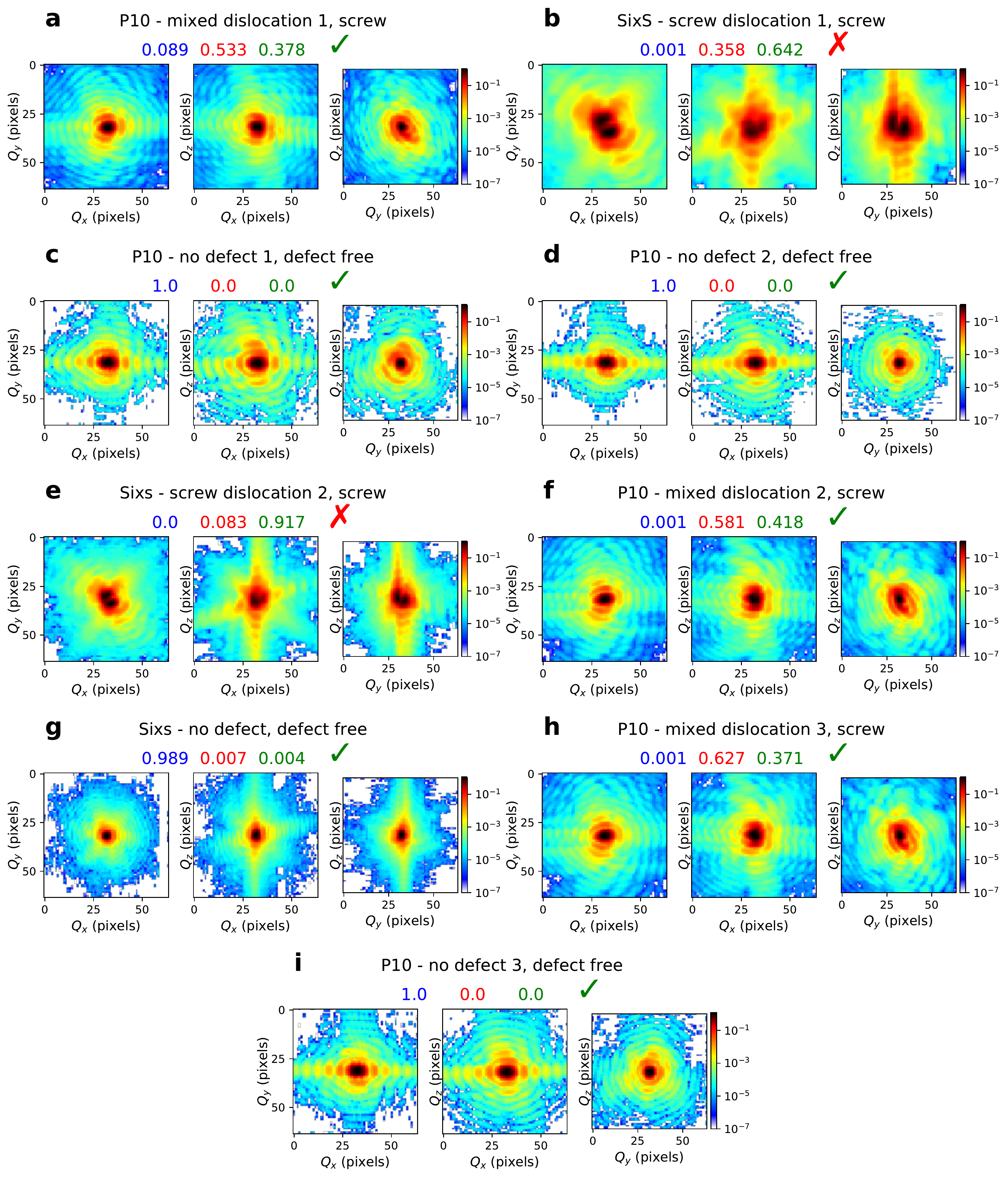}
\caption{\textbf{Predictions on the experimental data using a model trained on a 100\% RPD simulated dataset with a single atomic element (Au) and a large crystal size}. The predicted probability for each class is shown above each example where blue, red and green correspond respectively to defect free, screw and edge class. A green check mark/red cross in the title means that the prediction is correct/wrong.}
\label{fig:Result_single_element_random_dislo_position}
\end{figure}

\begin{figure}[ht!]
\centering
\includegraphics[width=.75\textwidth,angle=0]{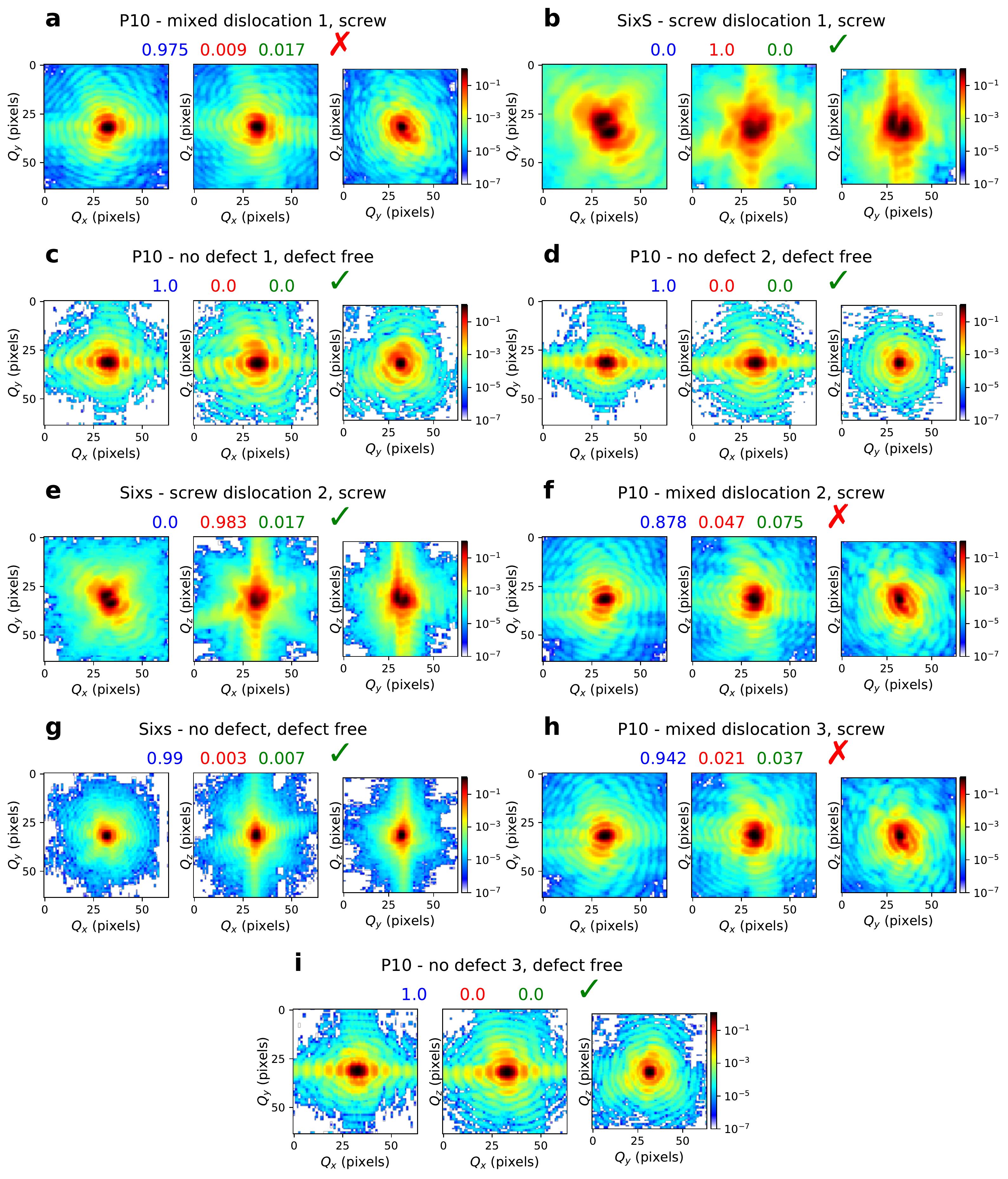}
\caption{\textbf{Predictions on the experimental data using a model trained on a 100\% CD simulated dataset with multiple atomic elements (Au, Pt) and a large crystal size}. The predicted probability for each class is shown above each example where blue, red and green correspond respectively to defect free, screw and edge class. A green check mark/red cross in the title means that the prediction is correct/wrong.}
\label{fig:Result_multi_element_dislo_near_center}
\end{figure}

Supplementary Figure \ref{fig:Result_single_element_random_dislo_position} shows the experimental predictions of a model trained on a dataset with a single atomic element (Au), random initial dislocation positions (100\% RPD dataset) and a large crystal size. While most predictions are correct, the SixS-screw dislocation is systematically predicted as an edge. This can be understood from the composition of the training dataset. Dislocations close to the free surfaces are frequently encountered in 100\% RPD training datasets. Upon relaxation, these dislocations tend to lose their edge or screw character and relax into a mixed dislocation (Supplementary Figure \ref{fig:edge_relax}). A training dataset with a  significant proportion of mixed dislocations, \textit{i.e.} where the edge or screw character of the dislocation is not well defined, can have detrimental consequences on the accuracy of the model, especially regarding the distinction between edge and screw defects. Therefore, we suggest that the the misclassification of SixS-screw dislocation into the edge class is linked to the  significant proportion of mixed dislocation in the 100\% RPD dataset.

If models trained on 100\% RPD datasets typically fail to predict correctly the edge or screw character, a training dataset with dislocations introduced only in the vicinity of the center of the nanocrystal (100\% CD datasets) is not without flaws. This is illustrated in Supplementary Figure \ref{fig:Result_multi_element_dislo_near_center}. which shows the experimental predictions of a 100\% CD dataset with multiple atomic elements (Au, Pt) and a large crystal size. In contrast with the model trained on a 100\% RPD dataset, the model trained on the 100\% CD dataset gives excellent predictions on the defect free and SixS-screw data, but incorrect predictions on the P10-mixed data, which is systematically identified as a defect free crystal. The reason behind these incorrect predictions is illustrated in Fig. 3 of the manuscript. As shown in Fig. 3d, the P10-mixed dislocation is close to the free surfaces, and yields much weaker distortions in the CXDPs than the centered dislocations of the simulated training dataset. We can infer that the model never learnt to identify the rather weak signature of an off-centered dislocation since it was only exposed to CD configurations, which create large distortions in the CXDPs. The lack of diversity of dislocation configurations/positions in the simulated training dataset results in a model that can predict accurately only specific dislocation configurations, \textit{i. e.} close to the center of the nanocrystal, when applied to experimental data. As a consequence, the P10 mixed dislocation is incorrectly identified  as a defect free crystal.


\subsection{Overview of the trainings performed on relaxed configurations}

Supplementary Tables \ref{table:tableS1} - \ref{table:tableS8} give an overview of the training results for the two crystal sizes and reveal several interesting trends. First, the accuracy of the predictions on the simulated test datasets is systematically very high (>86\%) (see Supplementary Tables \ref{table:tableS1} \& \ref{table:tableS5} for the training on the small and large crystal size, respectively). Interestingly it slightly decreases with an increasing fraction of RPD configurations in the training dataset. For instance the accuracy drops from 96.4\% for the 100\% CD datasets to 86.8\% for the 100\% RPD (small crystal size and Pt element in Supplementary Table \ref{table:tableS1}).
Training on the large crystal size also generally results in a slightly higher accuracy, but the difference is not significant (96.1\% overall accuracy in Supplementary Table \ref{table:tableS5} \textit{vs} 94.7\% accuracy for the small crystal size in Supplementary Table \ref{table:tableS1}).

Second, training on datasets with 100\% of CD systematically results in 100\% correct predictions for the experimental defect free and SixS-screw datasets (Supplementary Tables \ref{table:tableS2}, \ref{table:tableS3}, \ref{table:tableS6} \& \ref{table:tableS7})  but 0\% prediction for the P10-mixed dataset (Supplementary Tables \ref{table:tableS4} \& \ref{table:tableS8}), independently on the crystal size and the atomic element selected for the training dataset. This highlights the importance to add configurations with dislocations close to the free surfaces in order to increase the versatility of the model.

Third, training the model on datasets with 100\% RPD configurations results in an overall improvement of the predictions on the P10-mixed data. In particular the success rate dramatically improves for the Ag, Au and multi-elements datasets, but remains very low for Al and Pt datasets (Supplementary Tables \ref{table:tableS4} \& \ref{table:tableS8}). Moreover, this improvement comes at the expense of the accuracy on the SixS-screw dataset, in particular for the large crystal size (Supplementary Table \ref{table:tableS7}, 0\% of correct predictions). The success rate on the defect free configuration is also impacted, in particular for the small crystal size (Supplementary Table \ref{table:tableS2}, 45\% of correct predictions \textit{vs} 100\% for 100\% CD datasets). The only exception is the Au small crystal size datasets which retains a very high success rate for all the experimental datasets.
In order to improve the accuracy on the P10-mixed predictions while not impacting the success rate on the SixS-screw data, a better strategy consists in mixing CD and RPD datasets. For the large crystal configurations, a small fraction  of  RPD configurations is sufficient to dramatically improve the predictions on the P10-mixed data while retaining a very high success rate for the defect free and SixS-screw datasets (Supplementary Tables \ref{table:tableS6}-\ref{table:tableS8}). This is especially true for the multi-element Au-Pt dataset which performs very well when the ratio of RPD configurations in the dataset is between 20 and 35\%. This strategy is slightly less efficient for the small crystal datasets where the success rate is more element dependent (Supplementary Table \ref{table:tableS2}-\ref{table:tableS4}). Nonetheless, the introduction of a small fraction of RPD configurations in the training dataset is generally efficient to improve the success rate on the P10-mixed dataset, while keeping a high success rate on the defect free and SixS-screw experimental datasets (in particular for Ag, multi-element and to a lesser extent Pt training datasets). It also worth mentioning that this approach is not efficient on the Al dataset which performs poorly for the P10-mixed dataset, independently on the fraction of RPD configurations in the training dataset.

\begin{table}[ht!]
\small
\begin{tabular}{ || c | c | c | c | c | c | c || } 
\hline
\multicolumn{7}{||c||}{Accuracy (\%)} \\ 
\hline
& Ag & Al & Au & Pt & Ag-Al & Average \\
&    &    &    &    & -Au-Pt & \\        
\hline
100\% CD & 97.4 & 97.3 & 95.6 & 96.4 & 97.2 & 96.8 \\ 
\hline
87.5\% CD / 12.5\% RPD &  &  &  & 97.1 &  & 97.1 \\ 
\hline
75\% CD / 25\% RPD & 97.3 & 95 & 93.3 & 97 & 97.6 & 96 \\ 
\hline
62.5\% CD / 37.5\% RPD &  &  &  & 92.6 &  & 92.6 \\ 
\hline
50\% CD / 50\% RPD & 96.5 & 93.9 & 95.4 & 95.2 & 96.6 & 95.5 \\
\hline
25\% CD / 75\% RPD & 94.4 & 94 & 93.6 & 90.6 & 96.5 & 93.8 \\
\hline
12.5\% CD / 87.5\% RPD &  &  & 94.2 &  &  & 94.2 \\  
\hline
100\% RPD & 94.7 & 89.5 & 89.8 & 86.8 & 96.5 & 91.5 \\
\hline
Average & 96.1 & 93.9 & 93.7 & 93.7 & 96.9 & 94.7 \\
\hline
\end{tabular}
\caption{\textbf{Overview of the accuracy on the test dataset for the training performed on the small crystal size.}}
\label{table:tableS1}
\end{table}

\begin{table}[ht!]
\small
\begin{tabular}{ || c | c | c | c | c | c | c || } 
\hline
\multicolumn{7}{||c||}{Defect free crystals - correct predictions (\%)} \\ 
\hline
& Ag & Al & Au & Pt & Ag-Al & Average \\
&    &    &    &    & -Au-Pt & \\        
\hline
100\% CD & 100 & 100 & 100 & 100 & 100 & 100 \\ 
\hline
87.5\% CD / 12.5\% RPD &  &  &  & 100 &  & 100 \\ 
\hline
75\% CD / 25\% RPD & 50 & 75 & 100 & 75 & 100 & 80 \\ 
\hline
62.5\% CD / 37.5\% RPD &  &  &  & 100 &  & 100 \\ 
\hline
50\% CD / 50\% RPD & 25 & 50 & 100 & 50 & 75 & 60 \\
\hline
25\% CD / 75\% RPD & 50 & 0 & 50 & 50 & 75 & 45 \\
\hline
12.5\% CD / 87.5\% RPD &  &  & 75 &  &  & 75 \\  
\hline
100\% RPD & 25 & 25 & 100 & 25 & 50 & 45 \\
\hline
Average & 50 & 50 & 87.5 & 71.4 & 80 & 75.6 \\
\hline
\end{tabular}
\caption{\textbf{Percentage of correct predictions on the defect free crystals, small crystal size.}}
\label{table:tableS2}
\end{table}

\begin{table}[ht!]
\small
\begin{tabular}{ || c | c | c | c | c | c | c || } 
\hline
\multicolumn{7}{||c||}{SIXS screw - correct predictions (\%)} \\ 
\hline
& Ag & Al & Au & Pt & Ag-Al & Average \\
&    &    &    &    & -Au-Pt & \\        
\hline
100\% CD & 100 & 100 & 100 & 100 & 50 & 90 \\ 
\hline
87.5\% CD / 12.5\% RPD &  &  &  & 50 &  & 50 \\ 
\hline
75\% CD / 25\% RPD & 100 & 100 & 50 & 0 & 50 & 60 \\ 
\hline
62.5\% CD / 37.5\% RPD &  &  &  & 0 &  & 0 \\ 
\hline
50\% CD / 50\% RPD & 50 & 50 & 50 & 0 & 0 & 30 \\
\hline
25\% CD / 75\% RPD & 50 & 0 & 50 & 50 & 0 & 30 \\
\hline
12.5\% CD / 87.5\% RPD &  &  & 100 &  &  & 100 \\  
\hline
100\% RPD & 50 & 0 & 100 & 0 & 50 & 40 \\
\hline
Average & 70 & 50 & 75 & 28.6 & 30 & 50 \\
\hline
\end{tabular}
\caption{\textbf{Percentage of correct predictions for the SixS-screw data, small crystal size.}}
\label{table:tableS3}
\end{table}

\begin{table}[ht!]
\small
\begin{tabular}{ || c | c | c | c | c | c | c || } 
\hline
\multicolumn{7}{||c||}{P10 mixed - correct predictions (\%)} \\ 
\hline
& Ag & Al & Au & Pt & Ag-Al & Average \\
&    &    &    &    & -Au-Pt & \\        
\hline
100\% CD & 0 & 0 & 0 & 0 & 0 & 0 \\ 
\hline
87.5\% CD / 12.5\% RPD &  &  &  & 66 &  & 66 \\ 
\hline
75\% CD / 25\% RPD & 100 & 0 & 66 & 100 & 66 & 66.4 \\ 
\hline
62.5\% CD / 37.5\% RPD &  &  &  & 66 &  & 66 \\ 
\hline
50\% CD / 50\% RPD & 100 & 0 & 33 & 100 & 100 & 66.6 \\
\hline
25\% CD / 75\% RPD & 100 & 66 & 100 & 33 & 100 & 79.8 \\
\hline
12.5\% CD / 87.5\% RPD &  &  & 66 &  &  & 66 \\  
\hline
100\% RPD & 100 & 0 & 100 & 33 & 100 & 66.6 \\
\hline
Average & 80 & 13.2 & 60.8 & 56.9 & 73.2 & 59.7 \\
\hline
\end{tabular}
\caption{\textbf{Percentage of correct predictions for the P10 mixed data, small crystal size.}}
\label{table:tableS4}
\end{table}

\begin{table}[ht!]
\small
\begin{tabular}{ || c | c | c | c | c || } 
\hline
\multicolumn{5}{||c||}{Accuracy(\%)} \\ 
\hline
& Au & Pt & Au-Pt & Average \\
\hline
100\% CD & 96.1 & 97.6 & 98.1 & 97.3  \\ 
\hline
87.5\% CD / 12.5\% RPD  &  &  & 98 & 98 \\ 
\hline
80\% CD / 20\% RPD &  &  & 97.9 & 97.9\\  
\hline
75\% CD / 25\% RPD & 95.1 & 96.2 & 97.6 & 96.3 \\ 
\hline
62.5\% CD / 37.5\% RPD &  &  & 97.1 & 97.1 \\ 
\hline
50\% CD / 50\% RPD & 95.3 & 95.4 & 96.1 & 95.6  \\
\hline
25\% CD / 75\% RPD & 92.8 & 93.4 & 96.4 & 94.2 \\
\hline
100\% RPD & 90.5 & 91.2 & 94.9 & 92.2 \\
\hline
Average & 94 & 94.8 & 97.0 & 96.1 \\
\hline
\end{tabular}
\caption{\textbf{Overview of the accuracy on the test dataset for the training performed on the large crystal size.}}
\label{table:tableS5}
\end{table}

\begin{table}[ht!]
\small
\begin{tabular}{ || c | c | c | c | c || } 
\hline
\multicolumn{5}{||c||}{Defect free crystals - correct predictions (\%)} \\ 
\hline
& Au & Pt & Au-Pt & Average \\
\hline
100\% CD & 100 & 100 & 100 & 100  \\ 
\hline
87.5\% CD / 12.5\% RPD  &  &  & 100 & 100 \\ 
\hline
80\% CD / 20\% RPD &  &  & 100 & 100 \\  
\hline
75\% CD / 25\% RPD & 100 & 75 & 100 & 91.7 \\ 
\hline
62.5\% CD / 37.5\% RPD &  &  & 100 & 100 \\ 
\hline
50\% CD / 50\% RPD & 100 & 100 & 100 & 100  \\
\hline
25\% CD / 75\% RPD & 100 & 100 & 100 & 100 \\
\hline
100\% RPD & 100 & 75 & 75 & 83.3 \\
\hline
Average & 100 & 90 & 96.9 & 96.9 \\
\hline
\end{tabular}
\caption{\textbf{Percentage of correct predictions on the defect free crystals, large crystal size.}}
\label{table:tableS6}
\end{table}

\begin{table}[ht!]
\small
\begin{tabular}{ || c | c | c | c | c || } 
\hline
\multicolumn{5}{||c||}{SIXS-screw - correct predictions (\%)} \\ 
\hline
& Au & Pt & Au-Pt & Average \\
\hline
100\% CD & 100 & 100 & 100 & 100  \\ 
\hline
87.5\% CD / 12.5\% RPD  &  &  & 100 & 100 \\ 
\hline
80\% CD / 20\% RPD &  &  & 100 & 100 \\  
\hline
75\% CD / 25\% RPD & 0 & 50 & 100 & 50 \\ 
\hline
62.5\% CD / 37.5\% RPD &  &  & 0 & 0 \\ 
\hline
50\% CD / 50\% RPD & 0 & 50 & 50 & 33.3 \\
\hline
25\% CD / 75\% RPD & 0 & 50 & 0 & 16.7 \\
\hline
100\% RPD & 0 & 0 & 0 & 0 \\
\hline
Average & 20 & 50 & 56.3 & 50 \\
\hline
\end{tabular}
\caption{\textbf{Percentage of correct predictions for the SIXS-screw data, large crystal size.}}
\label{table:tableS7}
\end{table}

\begin{table}[ht!]
\small
\begin{tabular}{ || c | c | c | c | c || } 
\hline
\multicolumn{5}{||c||}{P10-mixed - correct predictions (\%)} \\ 
\hline
& Au & Pt & Au-Pt & Average \\
\hline
100\% CD & 0 & 0 & 0 & 0  \\ 
\hline
87.5\% CD / 12.5\% RPD  &  &  & 66 & 66 \\ 
\hline
80\% CD / 20\% RPD &  &  & 100 & 100 \\  
\hline
75\% CD / 25\% RPD & 0 & 66 & 100 & 55.3 \\ 
\hline
62.5\% CD / 37.5\% RPD &  &  & 100 & 100 \\ 
\hline
50\% CD / 50\% RPD & 100 & 0 & 66 & 55.3 \\
\hline
25\% CD / 75\% RPD & 66 & 0 & 100 & 55.3 \\
\hline
100\% RPD & 100 & 33 & 100 & 77.7 \\
\hline
Average & 53.2 & 19.8 & 79 & 63.7 \\
\hline
\end{tabular}
\caption{\textbf{Percentage of correct predictions for the P10-mixed data, large crystal size.}}
\label{table:tableS8}
\end{table}

\clearpage

Finally, while the overall success rate on the defective experimental datasets appears mostly unaffected by the crystal size, the introduction of RPD configurations in the training dataset has a much larger impact for the small crystal size than for the large crystal size. The overall success rate for the defect free configurations is therefore significantly higher for the large crystal size than for the small crystal size, and overall the distinction between defect free and defective crystals is more robust for the former (96.9\% accuracy) than for the latter (75.6\% accuracy), Supplementary Tables \ref{table:tableS1} \& \ref{table:tableS5}). 
Overall, adding diversity in the model (larger range of dislocation positions and/or atomic elements) is an efficient strategy in order to improve the robustness and accuracy of the CNN against a wide range of experimental configurations.


\subsection{2 classes model}

While the 3 classes neural network model presented in the manuscript can make very accurate predictions on experimental data, the confusion matrix for the simulated data shows that some screw dislocations are predicted as edge and \textit{vice versa}. On the other hand, almost all defect free crystals are predicted in the correct class. This implies that a 2 classes model predicting either a defect free crystal or one with a defect could be even more accurate.
The architecture of this simpler 2 classes model is shown in Supplementary Figure \ref{fig:2classes_architecture}. Again, 3D CXDPs are used as input data for the CNN and encoded through 3 convolution layers and 2 fully connected layers to predict the  probabilities of the 2 classes (defective or defect free crystal).

\begin{figure}[ht!]
\centering
\includegraphics[scale = 0.7]{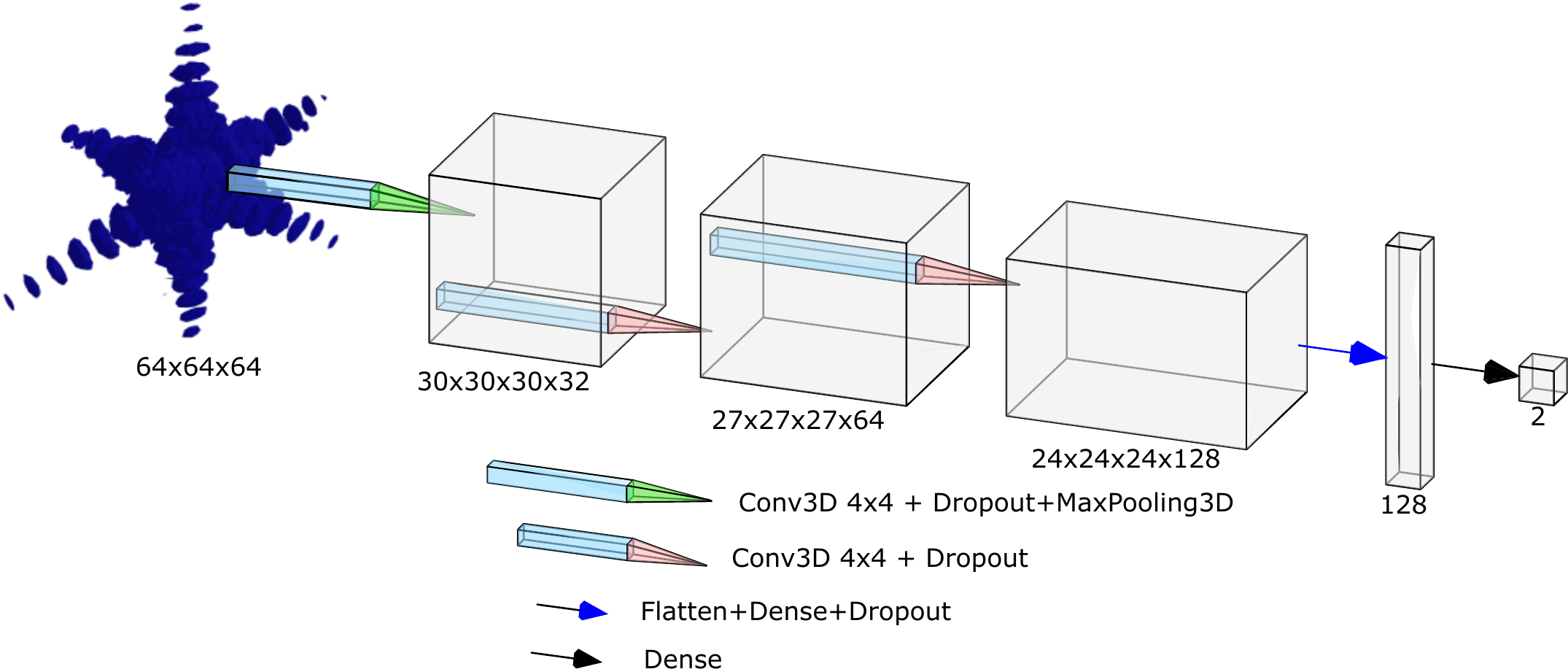}
\caption{\textbf{Simplified neural network architecture used for the 2 classes model}. The last  fully connected layer returns 2 probability values (defect free or defective crystal). }
\label{fig:2classes_architecture}
\end{figure}

\begin{figure}[ht!]
\centering
\includegraphics[scale = 0.3]{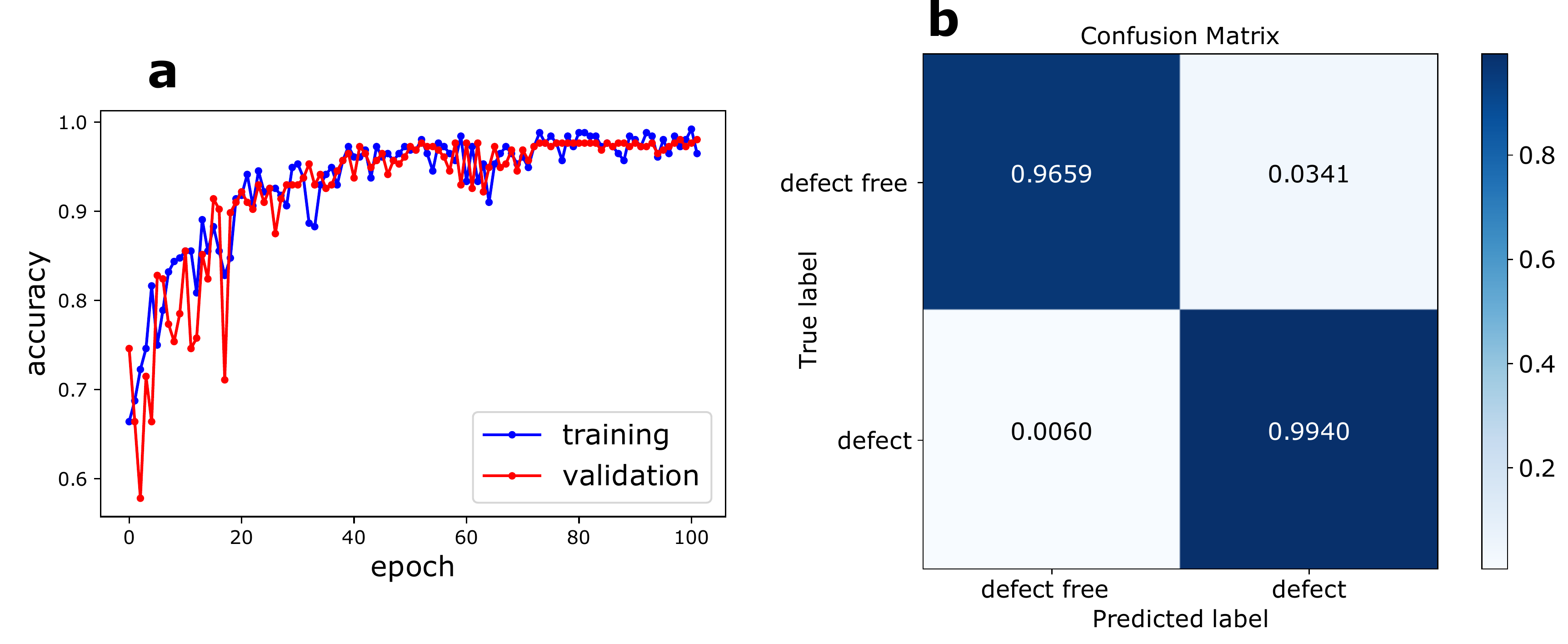}
\caption{\textbf{Training and performances of the 2 classes model}. \textbf{a} Accuracy for the training and validation set during the model training. \textbf{b} Confusion matrix for the 2 classes model calculated with the simulated examples test set.}
\label{fig:2classes_confusion}
\end{figure}

\begin{figure}[ht!]
\centering
\includegraphics[width=.63\textwidth,angle=0]{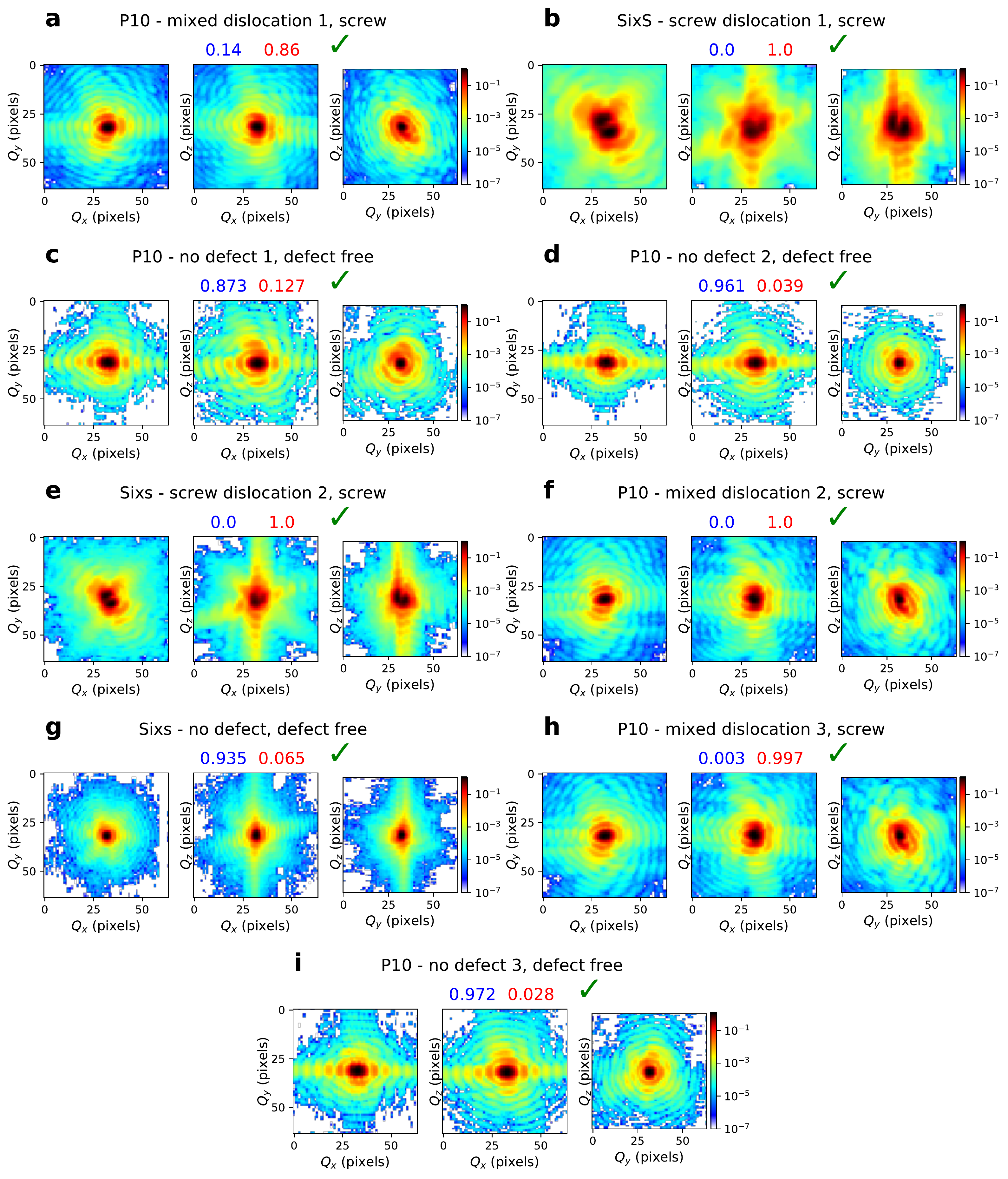}
\caption{\textbf{Predictions of the 2 classes model on the experimental data.} For each example, the predicted probabilities for the perfect and defect classes are shown respectively in blue and red in the title. All predictions are correct, each of them with a very high probability for the correct class.}
\label{fig:2classes_results}
\end{figure}

The model trained on 4300 simulated examples reaches an accuracy of 98.44\% on the test dataset of 256 examples. The model is trained with an Adam optimizer using a learning rate of $2\times10^{-3}$ and a batch-size of 32. In order to fine-tune the model, we decrease the learning-rate and increase the batch-size after few epochs. This is visible in the accuracy of the training and validation sets shown in Supplementary Figure \ref{fig:2classes_confusion}a where the oscillations become very small by the end of the training procedure. The confusion matrix of the test dataset is shown in Supplementary Figure \ref{fig:2classes_confusion}b and both non-diagonal elements are very small, thus indicating a very high score on each class.

Supplementary Figure \ref{fig:2classes_results} shows the  prediction of the 2 classes model on the experimental data. All predictions are correct with slightly higher probabilities than in the 3 classes model, the lowest one being for the "P10 - mixed dislocation" with an 86\% probability of a defective crystal.


\subsection{Occlusion sensitivity test}

The occlusion sensitivity test is a good method to check which features/regions of the the 3D CXDP are used by the neural network model to calculate the class probabilities \cite{zeiler2014visualizing}. We selected a simulated 3D CXDP of a crystal containing a screw dislocation shown in Supplementary Figure \ref{fig:center prediction}a. The model gives a correct prediction with a 87\% screw class probability. A region of the 3D array is then masked with a sphere of 13 pixels radius at a given position in the 64x64x64 diffraction array as shown in Supplementary Figure \ref{fig:center prediction}b. The trained model is subsequently used on the masked array to compute a new prediction of the screw probability. Finally, the predictions are calculated for all the mask locations represented in Supplementary Figure \ref{fig:center prediction}c, where the dots correspond to the mask center position and the screw probability is shown in a color-scale. We observe that the probability drops drastically when the mask is close to the center of the 3D array suggesting that the most important information to achieve accurate predictions is in the vicinity of the Bragg peak.

\begin{figure}[ht!]
\centering
\includegraphics[width=0.8\textwidth, angle=0]{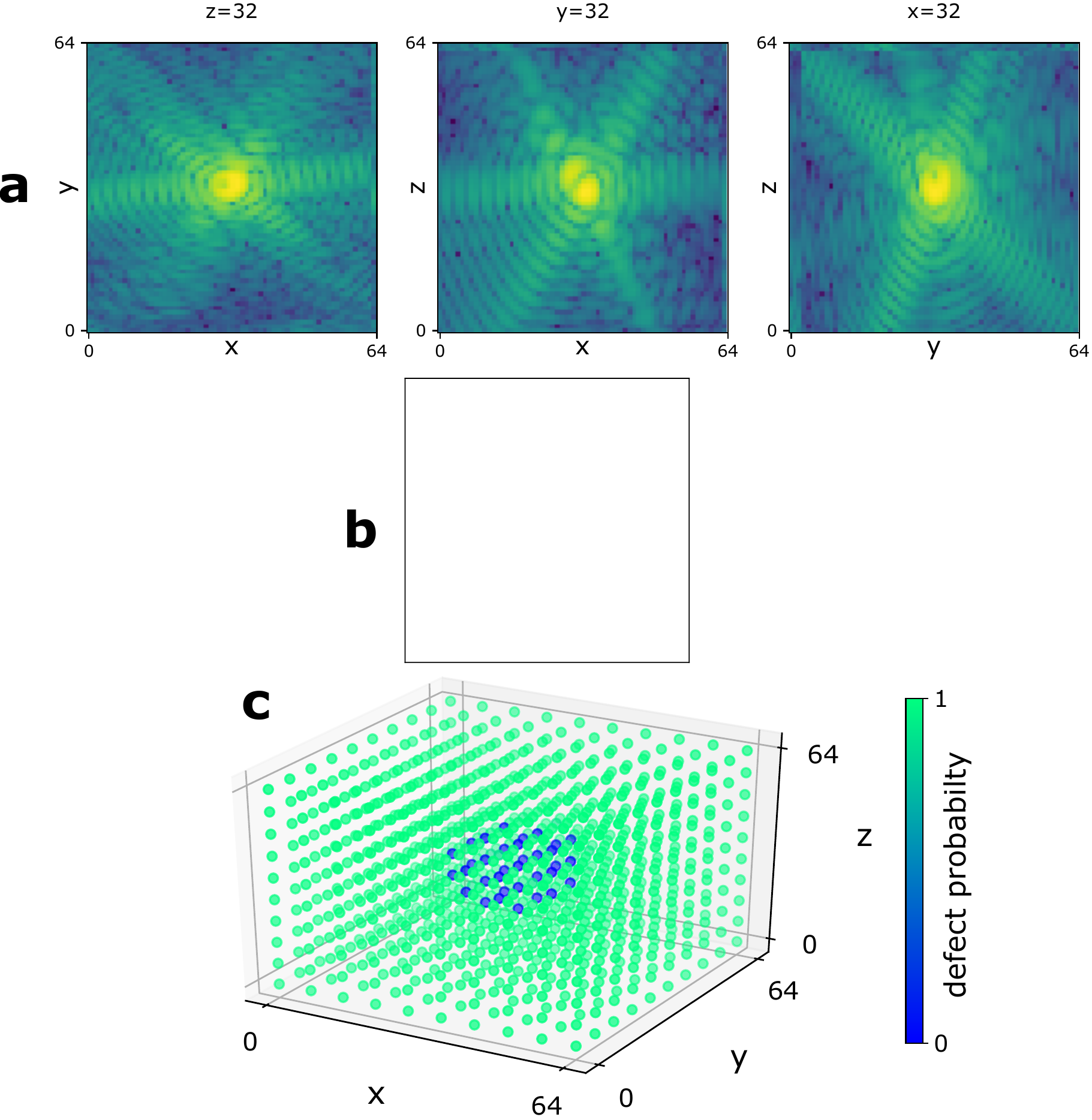}
\caption{\textbf{Occlusion sensitivity test}. \textbf{a} 2D slices in log-scale of the 3D simulated example of a screw dislocation used to test the occlusion sensitivity of the neural network. \textbf{b} 2D representation of the circular mask. The mask center is depicted as a white cross. \textbf{c} Model prediction of the screw probability as a function of the mask center position. The probability drops drastically when the mask is close to the center of the diffraction pattern.}
\label{fig:center prediction}
\end{figure}

\clearpage


\subsection{Influence of rotation and zoom on the CNN predictions}

\begin{figure}[ht!]
\centering
\includegraphics[width=0.85\textwidth, angle=0]{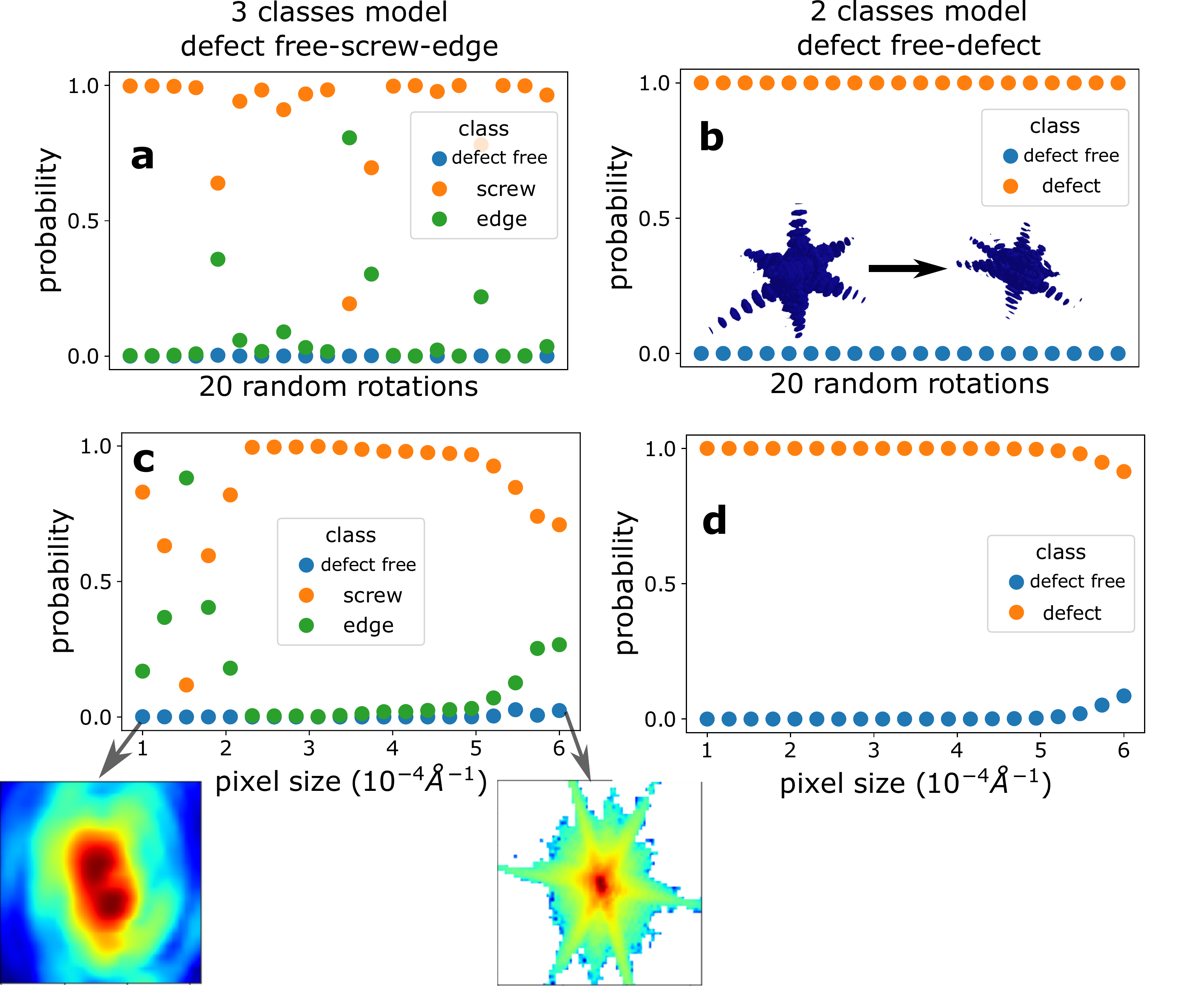}
\caption{\textbf{Influence of rotation and zoom on the CNN predictions}. \textbf{a} 3 classes model prediction for 20 random rotations of an experimental screw example. \textbf{b} Same with a 2 classes model prediction. \textbf{c} 3 classes model prediction for different pixel sizes. A 2D slice of the diffraction pattern for the smallest and largest pixel sizes are shown below. \textbf{d} Same with a 2 classes model prediction.}
\label{fig:rotation zoom}
\end{figure}

In order to evaluate the influence of the crystal orientation on the model predictions, we rotated an experimental diffraction pattern randomly with a combination of 3 rotations along the x, y and z axes as shown in the inset of Supplementary Figure \ref{fig:rotation zoom}b. To avoid any artefact in the rotated examples, we used a 3D interpolation of the unrotated diffraction. 

The 3 classes model predictions for 20 random rotations are shown in Supplementary Figure \ref{fig:rotation zoom}a. Despite some variations, we observe that most examples are correctly predicted  as a screw dislocation. One of them is predicted as an edge meaning that our model is slightly dependent on the crystal orientation. For a larger statistic, we make a prediction on 200 random rotations and observe that 93\% of them are predicted as screw, meaning that our model is mostly independent on the crystal orientation. 

From Supplementary Figure \ref{fig:rotation zoom}a, it can also be seen that none of the examples are predicted as a defect free crystal, demonstrating that the model can always predict the defective character of the nanocrystal. To verify this, we trained the 2 classes model shown in Supplementary Figure \ref{fig:2classes_architecture}, predicting either a defect free or defective crystal. We used again the experimental screw example with 20 random rotations and calculated the predictions shown in Supplementary Figure \ref{fig:rotation zoom}b. As expected, all predictions are correct for this model, meaning that they are independent of the crystal orientation.

As a second test, we created several examples with different pixel sizes \textit{i.e.} resolution or reciprocal space sampling using the same 3D interpolated experimental example with a screw dislocation. Our 3 classes model prediction is calculated for each resolution as shown in Supplementary Figure \ref{fig:rotation zoom}c where a 2D slice of the diffraction is shown for the highest and lowest resolution. The model prediction is worse for very small and very large pixels but always correct between these 2 extremes. 
Again, we observe in the small pixel regime that our model fails to predict the edge or screw character of the dislocation (even though it still makes the correct screw prediction for most pixel sizes). The 2 classes model (Supplementary Figure \ref{fig:rotation zoom}d) is even less sensitive to the pixel size and always correctly predict the defective character of the nanocrystal. In conclusion, the 3 classes model can generalize to very diverse experimental measurements and the simple 2 classes model is even more robust to these variations.


\subsection{Maximum intensity plots}

\begin{figure}[ht!]
\centering
\includegraphics[width=0.99\textwidth, angle=0]{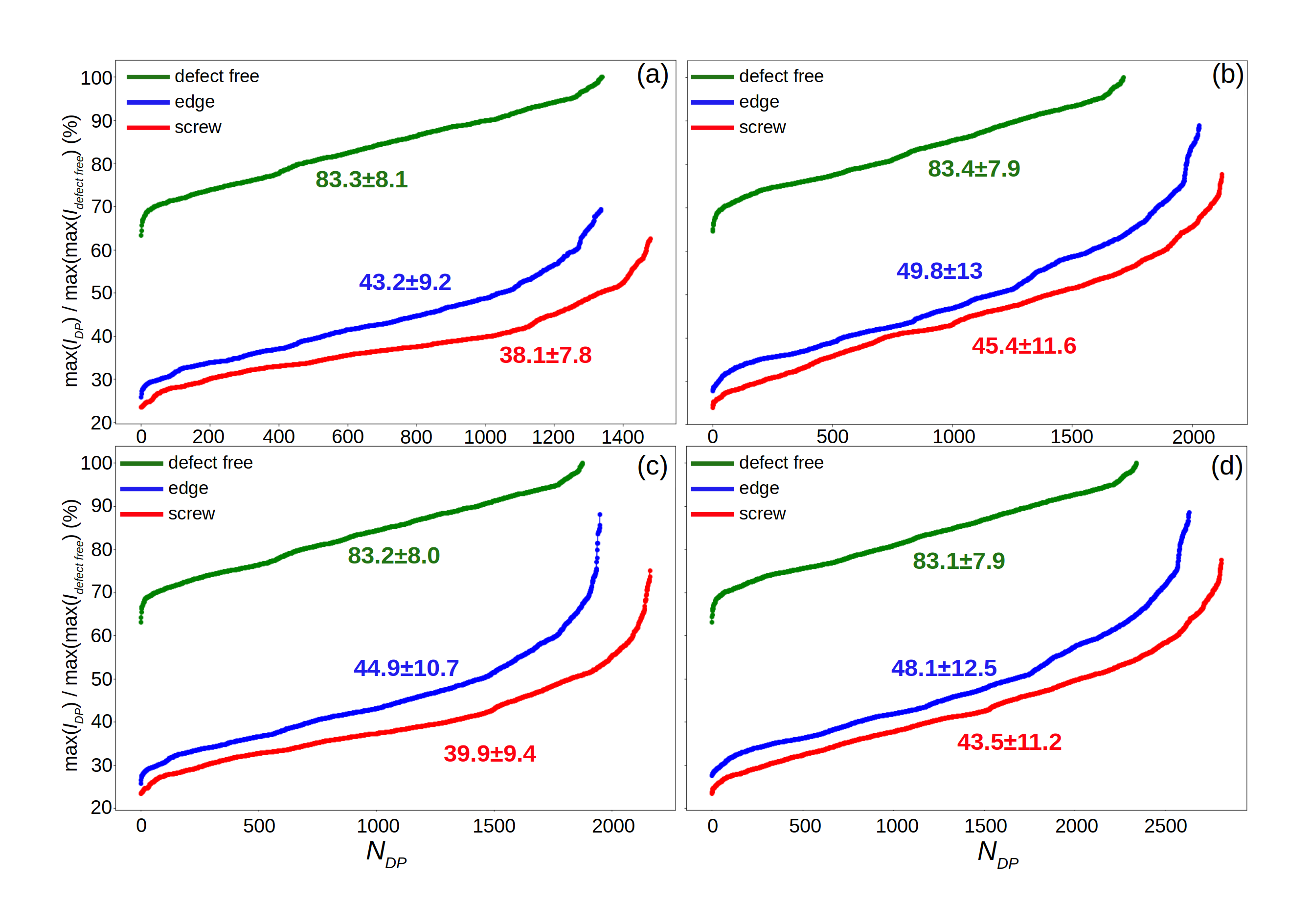}
\caption{\textbf{Maximum of intensity of the simulated diffraction patterns (DPs) with respect to the maximum of intensity of a defect free configuration with a similar number of atoms.} \textbf{a} Training dataset with 100\% CD configurations, \textbf{b} 100\% RPD, \textbf{c} 75\% CD / 25\% RPD and \textbf{d} 25\% CD / 75\% RPD. The DPs were calculated from Au small crystals configurations. The average of the maxima of intensity together with their standard deviation are indicated on the plots for each type of configuration.}
\label{fig:max_intensity}
\end{figure}

We present in this section a quick method to estimate the distribution of dislocation positions in the simulated nanocrystals. It is based on the calculation of the maximum of the intensity scattered by a given atomistic configuration  with respect to the maximum of the intensity scattered by a defect free crystal with a similar number of atoms. This quantity is largely affected by the dislocation position in the crystal. Dislocations introduced close to the center of the nanocrystal, induce large distortions in the vicinity of the Bragg peak, accompanied by a large decrease of the maximum intensity (corresponding typically to 20-35\% of the maximum intensity of a defect free crystal with a similar number of atoms for a \{1 1 1\}-type reflection \cite{dupraz2015signature}). The distortions and the corresponding decrease of the maximum intensity are much less pronounced if the dislocation is introduced close to the free surfaces. For such configurations, the maximum of intensity corresponds typically to more than 60\% of the maximum of a defect free crystal. 
The correlation between the distribution of the maxima of intensity and the ratio of CD and RPD configurations in the training dataset is illustrated in Supplementary Figure \ref{fig:max_intensity}. 
For a 100\% CD training dataset (Supplementary Figure \ref{fig:max_intensity}a), all dislocations are introduced close to the center, within a range not exceeding 10\% of the lateral size of the nanocrystal. As a consequence, they yield a significant decrease of the maximum intensity for both screw and edge configurations: $\langle max(I_{screw})\rangle $ = 38.1 $\pm$ 7.8, $\langle max(I_{edge})\rangle$ = 43.2 $\pm$ 9.2.
Note that the maximum of intensity also exhibits some variations for the defect free configurations. This is mostly due to the fact that the number of atoms is not constant (100000-140000 atoms and 800000-950000 atoms for the small and big nanocrystals, respectively). Indeed the (1 1 1) interface plane is cut at a random height in a range corresponding to 65\% to 75\% of the height of a free standing Wulff particle. In addition, changes in $\delta q$, which is also randomized, can result in slight variations of the maximum of intensity.

The larger range of dislocation positions allowed for the RPD datasets is reflected in the distribution of the maxima of intensity of the 100\% RPD dataset. As shown in Supplementary Figure \ref{fig:max_intensity}b, the average value of the maxima increases significantly for both screw and edge defects : $\langle max(I_{screw})\rangle$ = 45.4 $\pm$ 11.6, $\langle max(I_{edge})\rangle$ = 49.8 $\pm$ 13 . 
The dislocations introduced close to the free surfaces induce a limited decrease of \textit{max(I)}. For a small fraction of the defective crystals, \textit{max(I)} is even within the range of the maxima of intensity computed for defect free configurations  (\textit{max(I)} > 65\%). The larger range of dislocation positions therefore results in a wider distribution of the maxima of intensity, hence the increase of the standard deviation together with the increase of the average value. 
The last two remaining datasets shown in Supplementary Figure \ref{fig:max_intensity} can be seen as intermediate cases between the 100\% CD and 100\% RPD datasets. The maximum intensity distribution in the 75\% CD / 25\% RPD dataset (Supplementary Figure \ref{fig:max_intensity}c) is rather similar to the one in the 100\% CD. Contrary to the latter, a small fraction of the configurations in the dataset contains dislocations close to the free surfaces (\textit{max(I)} > 65\%). This small fraction is sufficient to yield an increase of both the average value and standard deviations of \textit{max(I$_{screw}$)} and \textit{max(I$_{edge}$)}. Finally, the distribution of the maxima of intensity in the 25\% CD / 75\% RPD dataset is similar to the one in the 100\% RPD dataset. The larger fraction of CD configurations in the dataset is reflected in the lower values of $\langle max(I_{screw})\rangle $, $\langle max(I_{edge})\rangle$ and of their standard deviations $\sigma$(\textit{max(I$_{screw}$)}), $\sigma$(\textit{max(I$_{edge}$)}). 
Overall, there is therefore a clear correlation between the fraction of RPD configurations in the dataset and the resulting distribution of the maxima of intensity. A higher fraction of RPD results in an increase of $\langle max(I_{dislo})\rangle$ and $\sigma$(\textit{max(I$_{dislo}$)}) , while a lower fraction results in a decrease of these two quantities.


\subsection{CNN training reproducibility}

When creating a CNN model with tensorflow, the weights of each convolutional and fully connected layers are initialized randomly. Moreover, the instances in the training dataset are processed in a random for each epoch, and the CNN training is stopped manually once the validation set accuracy reaches a maximum value. Therefore, the CNN output will be slightly different for each training, and one could ask if a CNN always gives a similar probability distribution for different training. In order to evaluate the similarity of the probability distributions, we trained 2 models with the same architecture. Identical training and validation datasets were used for both model. The confusion matrices calculated with the same test dataset are shown in Supplementary Figure \ref{fig:reproducibility}. The matrix element are very similar between model 1 and model 2 demonstrating the reproducibility of the CNN training.

\begin{figure}[ht!]
\centering
\includegraphics[width=0.99\textwidth, angle=0]{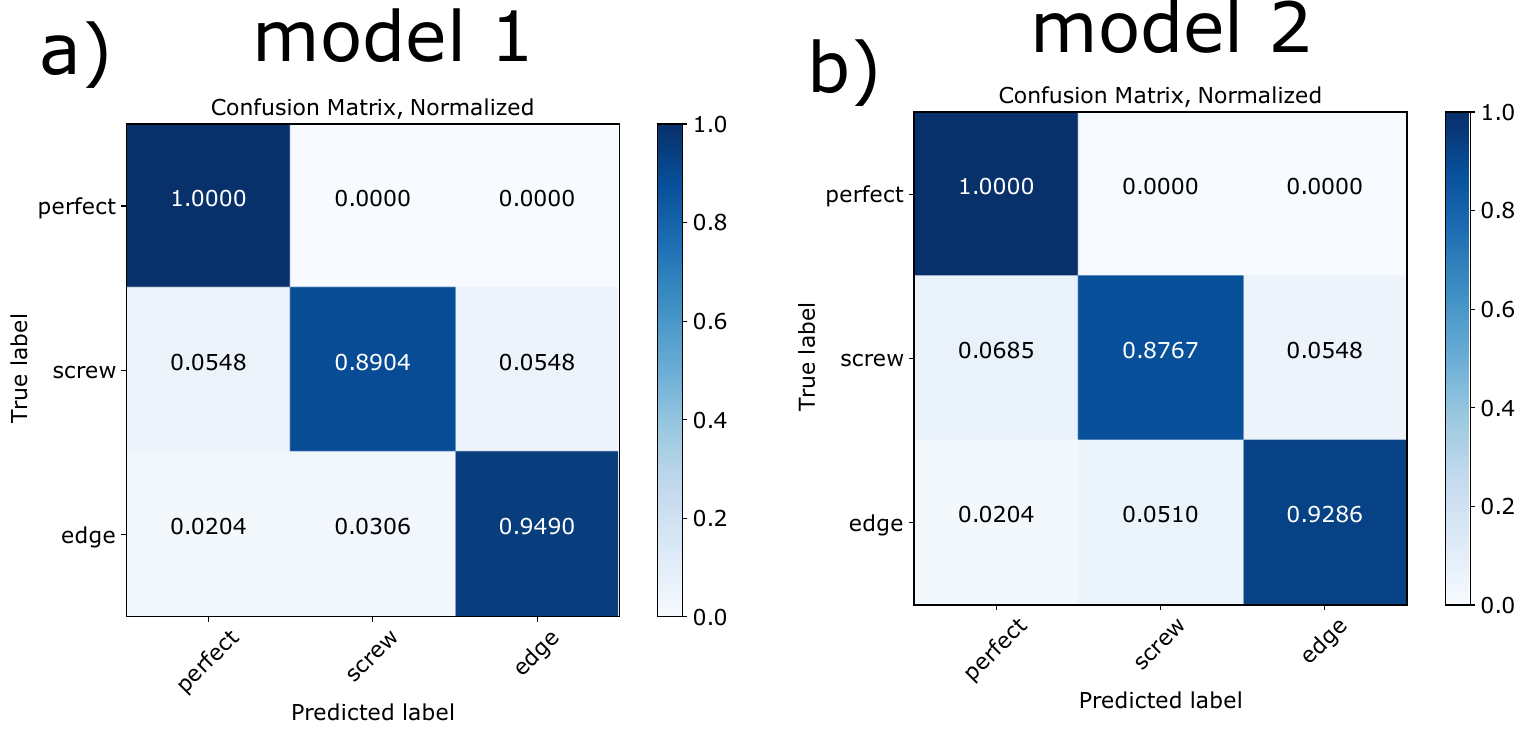}
\caption{\textbf{Confusion matrices for 2 CNN models with identical architecture and trained with the same dataset.} These matrices were both calculated with the same test dataset.}
\label{fig:reproducibility}
\end{figure}

\clearpage

\input{ms_sup.bbl}


%% file: ms_sup.bbl
%